\documentclass{jfm}

\usepackage{graphicx}
\usepackage{newtxtext}
\usepackage{newtxmath}
\usepackage{natbib}
\usepackage{hyperref}
\hypersetup{
    colorlinks = true,
    urlcolor   = blue,
    citecolor  = black,
}

\newcommand{\RomanNumeralCaps}[1]

\usepackage{rotating}
\usepackage{float}
\usepackage{array}
\usepackage[utf8]{inputenc}
\usepackage{braket}
\usepackage[ISO]{diffcoeff}
\usepackage{amsmath}
\usepackage{caption}
\DeclareCaptionFont{viii}{\fontsize{8}{10}\selectfont}
\usepackage{subfig}
\usepackage{bm}
\usepackage{mathbbol}
\usepackage{appendix}
\usepackage{multirow}

\newcommand{\D}[1]{\mathcal{D}_{#1}}
\newcommand{\Emin}{E_{\mathrm{min}}}
\newcommand{\alphamin}{\alpha_{\mathrm{min}}}
\newcommand{\tcrit}{t_{\mathrm{crit}}}

\title{Modelling flow-driven pore closure of weakening poroelastic media}

\author{Matthew V. Ghosh\aff{1}
  \corresp{\email{matthew.ghosh@maths.ox.ac.uk}},
  Matthew G. Hennessy\aff{2},
  Andreas Münch\aff{1} 
 \and Sarah L. Waters\aff{1}}

\affiliation{\aff{1}Mathematical Institute, University of Oxford, Radcliffe Observatory Quarter, Woodstock Road, Oxford, OX2 6GG, UK
\aff{2}School of Engineering Mathematics and Technology, University of Bristol, Ada Lovelace Building, Bristol, BS8 1TW, UK}

\begin{document}
\maketitle

\begin{abstract}
    Poroelastic materials, such as polymer tissue scaffolds, porous rocks, and hydrogels, can weaken due to interactions between the solid skeleton and chemical species in the interstitial fluid. We develop a mathematical model for a poroelastic material to provide fundamental mechanistic insight into how weakening the material can affect the time-varying mechanics of the system. Our model couples large-deformation poroelasticity with an advection-diffusion equation for the solute. Furthermore, we introduce a decay equation for the material stiffness, whose rate of decay depends on the solute concentration. In this way, we describe a three-way coupling between poroelastic deformation, weakening of the skeleton and transport of solute through the material. We exploit numerical and analytical techniques to reveal the flow-driven uniaxial compression of a weakening poroelastic material and determine parameter regimes for which weakening the material facilitates pore closure at the downstream boundary. We identify parameter regimes in which (1) a steady state is attained without pore closure, (2) pore closure occurs at a finite time or (3) the pores close instantaneously; we uncover case (2) through the introduction of weakening into the system. We provide insights into the relationship between the differing behaviours and the separation between the timescales of the system. For systems with slow weakening, we derive a leading-order approximation for the time of pore closure, treating the ratio of the timescales of poroelastic relaxation and weakening as a small parameter, and investigate the accuracy of this approximation and the new behaviours that arise when these timescales become comparable.
\end{abstract}




\section{Introduction} \label{sec:introduction}


Poroelasticity can be used to model both geological systems, such as porous rocks with large elastic moduli~\citep{Liang2008}, and biomedical materials, such as ureteric stents~\citep{Barros2016} and polymer scaffolds used in tissue repair~\citep{LeBlon2013}. Chemical species carried by the interstitial fluid can weaken these structures and, in some cases, cause the porous solid matrix to degrade. In this work, we use the term `weakening' to describe the reduction of material stiffness, whereas by `degradation' we mean material breaking off from the solid skeleton. Salts dissolved within the fluid can weaken and degrade rocks by crystallisation or hydration~\citep{Charola2000,Nooraiepour2025}. In contrast, the invasion of low-salinity fluids into porous rocks can induce the dissolution of the matrix into the fluid, which causes the material properties to evolve~\citep{Tan2023}.


Hydrogels, which are composed of cross-linked networks of polymer chains, and other soft poroelastic materials, can weaken upon interaction with certain chemical species, such as enzymes~\citep{Meyvis1999} and salts~\citep{Li2022} dissolved within the interstitial fluid, or simply by the solvent itself~\citep{Ozcelik2016}; this process is exploited in several medical applications. Each application benefits from an understanding of the precise mechanism and timescales of weakening and degradation of these materials. In pharmacokinetics, hydrogels can be used to control the concentration of a drug released into the bloodstream~\citep{Peppas2000}. For example, \citet{Meyvis1999} studied how dextran methacrylate (Dex-MA) hydrogels weakened and degraded in the presence of the enzyme dextranase; these hydrogels have been considered as constituents for drug-delivery systems~\citep{Petrovici2023}. Porous polymer scaffolds used in tissue engineering are designed such that as they biodegrade, new tissue grows in their place~\citep{Chen2001}. \citet{LeBlon2013} used a salt-leaching technique to create macropores in the polymers. The polymer scaffolds were then submerged in simulated body fluid (SBF) to assess the biodegradation of the scaffolds~\citep{Wierzbicka2023}. \citet{Barros2016} developed a pure hydrogel ureteric stent made of sodium alginate, which weakens and degrades when urine pumped through the ureter deposits salts on the device. 


Many of these biomedical devices are subject to a flow of fluid~\citep{Chen2006} or placed in a flow \textit{in vivo}, which leads to a coupling between two key physical processes that characterise Biot poroelasticity: pressure-driven flow and elastic deformation~\citep{Biot1941}. However, the inclusion of a chemical species that weakens the solid matrix introduces further couplings, meaning that a model going beyond that of Biot is required to capture the complex dynamics of the system.



Several physical mechanisms can induce weakening and degradation in hydrogels. Hydrolytic degradation has been modelled by assuming the density of crosslinks~\citep{Dhote2014} or the proportion of ester bonds in the gel~\citep{Pan2022} exponentially decay in time due to first-order chemical reactions. For enzymatic degradation, the rate of decay in cross-linking density can also depend proportionally on the concentration of enzyme present in the system~\citep{Vernerey2012,Dhote2014,AbiAkl2019}. Experimental work by \citet{Meyvis1999} showed that in some cases, the rate of decay of the elastic moduli is proportional to the concentration of enzyme. This relation gives rise to the possibility of spatially varying material properties due to a non-uniform distribution of enzymes. Spatially varying material properties due to inherent heterogeneities have also been modelled in poroelastic materials~\citep{Godard2025}.


One phenomenon that can arise from fluid-driven compression of poroelastic materials is pore closure, in which the local porosity of a deformable porous material decreases to zero. Pore closure can occur in systems with low initial porosity, high pH~\citep{Frendenberg2011} or large fluid pressure~\citep{David1993}. In addition, \citet{Schwartz2019} demonstrated numerically that in gas shales, softer pores experience more pore closure than stiffer ones. It follows that weakening the solid matrix can, over time, cause pore closure at a flow rate where it would not have occurred previously. \citet{MacMinn2016} noted that in the case of a non-weakening poroelastic material obeying Hencky elasticity, there is an upper limit to the pressure drop that can be applied across the material. For pressures above this upper limit, the pores close instantaneously at one of the boundaries. We hypothesise that weakening the material decreases the maximal pressure drop. In Biot poroelasticity, when pore closure occurs, the fluid volume fraction reaches zero and the model breaks down. In the work by~\citet{Jannesari2026}, the authors posited a new poroelastic model for pore closure. In their model, as the porosity approaches zero, the fluid velocity approaches the solid velocity, which allows them to model zero-porosity regions without the model breaking down.


In this work, we provide mechanistic understanding of both the dynamic and long-term behaviour of poroelastic materials weakening due to a solute, carried by the interstitial fluid, interacting with the solid skeleton. In particular, we use mathematical modelling techniques to investigate how weakening the stiffness of the material can, in some cases, lead to a localised closure of pores and the loss of a steady state. We investigate a canonical one-dimensional geometry for poroelastic deformation (see e.g.~\citep{MacMinn2016}) to illustrate the coupling between weakening, poromechanics and solute transport in a simple geometry that could be realised experimentally while retaining many key aspects of the physical examples detailed above. The system is characterised by four timescales, namely those associated with poroelastic relaxation, fluid advection, solute diffusion and weakening. We combine large deformation poroelasticity~\citep{Biot1941, MacMinn2016} with solute transport through a porous medium from~\citet{Fiori2025}. \citet{Fiori2025} did not suppose that the solute interacted with the solid skeleton, and to complete our model framework, we introduce a weakening law for the solid skeleton inspired by a model for the enzymatic degradation of hydrogels~\citep{AbiAkl2019}.

\citet{Chen2024} formulated a general three-dimensional model for the dissolution of a small-deformation linear poroelastic solid due to ionic species in the interstitial fluid in the context of geomaterials. The instantaneous bulk and shear stiffness of the deformable skeleton are proportional to the cube of the solid fraction and hence decrease as the skeleton dissolves. In contrast, in our model, we consider large-deformation nonlinear poroelasticity and pose an equation for the rate of weakening, which depends explicitly on the local concentration of solute.


In \S\ref{sec:3D-model}, we present a novel three-dimensional continuum model of the system. In \S\ref{sec:channel-model}, we consider the canonical example of flow-driven uniaxial compression of a poroelastic material. We nondimensionalise the governing equations, boundary and initial conditions, which reveals the four characteristic timescales of the system, and two experimentally motivated parameter sets are discussed. In \S\ref{sec:steady-state} we derive the steady state of the model. We then explore the long-term behaviour of systems with slow weakening in \S\ref{sec:slow-weakening}, taking the ratio of poroelastic relaxation to weakening timescales to be small, and use both analytical and numerical techniques to determine conditions under which pore closure occurs. We relax the slow weakening assumption in \S\ref{sec:comparable-timescales} and investigate how certain regions of parameter space give rise to inhomogeneous weakening of the material, in contrast to \S\ref{sec:slow-weakening}. We also determine the extent to which the conditions determined in \S\ref{sec:slow-weakening} for the existence of a steady state and pore closure hold without the slow weakening assumption.

\section{Three-dimensional continuum model} \label{sec:3D-model}

In this section, we detail the governing equations for a general three-dimensional poroelastic material undergoing deformations and weakening. We delay specification of boundary and initial conditions until \S\ref{sec:channel-model}, in which we present the canonical problem we consider. We suppose that an incompressible interstitial fluid flows between incompressible solid grains of the material and that the solid skeleton weakens in the presence of a chemical species dissolved in the interstitial fluid.

The formulation of the poroelastic material follows~\citet{MacMinn2016}. We model the material with Eulerian coordinates $\boldsymbol{x}$ fixed in space and Lagrangian coordinates $\boldsymbol{X}$ in the reference configuration. The reference configuration corresponds to a saturated poroelastic medium with uniform porosity $\phi_{f,0}$. We derive all equations in an Eulerian framework. In addition, we use the subscripts $s$ and $f$ to refer to the solid and fluid, respectively. Firstly, the volume fractions for the solid and fluid, $\phi_{s}$ and $\phi_{f}$ respectively, satisfy 

\begin{equation} \label{gen:solid-phi_s}
    \phi_{s} + \phi_{f} = 1, 
\end{equation}

\noindent assuming no voids. We define the solid displacement, $\boldsymbol{u}_{s}$, as

\begin{equation} \label{gen:solid-u_s}
    \boldsymbol{u}_{s}(\boldsymbol{x}, t) = \boldsymbol{x} - \boldsymbol{X}(\boldsymbol{x}, t).
\end{equation}

\noindent In this Eulerian framework, we define the deformation gradient tensor, $\mathsfbi{F}$, in terms of spatial gradients of $\boldsymbol{X}(\boldsymbol{x}, t)$ as follows:

\begin{equation} \label{gen:solid-F}
    \mathsfbi{F} = \left(\bm{\nabla}\boldsymbol{X}\right)^{-1} = \left(\mathsfbi{I} - \bm{\nabla}\boldsymbol{u}_{s}\right)^{-1},
\end{equation}

\noindent where we have used (\ref{gen:solid-u_s}) to express $\mathsfbi{F}$ in terms of the solid displacement, and $\mathsfbi{I}$ is the identity tensor. We denote the determinant of $\mathsfbi{F}$ by $J$, which measures the volume change of material elements relative to the initial reference state. As we assume that each individual solid grain is incompressible, compression of the material can only occur through removal of fluid from the pore space. Hence, $J$ is simply the ratio between the initial solid fraction and the current solid fraction:

\begin{equation} \label{gen:solid-J-phi}
    J = \frac{1 - \phi_{f,0}}{1 - \phi_{f}},
\end{equation}

\noindent where we have used (\ref{gen:solid-phi_s}) to eliminate the solid volume fraction. The solid velocity, $\boldsymbol{v}_{s}$, is defined as

\begin{equation} \label{gen:solid-v_s}
    \boldsymbol{v}_{s} = \mathsfbi{F}\frac{\p \boldsymbol{u}_{s}}{\p t}.
\end{equation}

\noindent Since solid grains are individually incompressible, we use (\ref{gen:solid-phi_s}) to express the conservation of mass equation for the solid phase as follows:

\begin{equation} \label{gen:solid-mass}
    \frac{\p\phi_{f}}{\p t} - \bm{\nabla}\boldsymbol{\cdot}\left((1 - \phi_{f})\boldsymbol{v}_{s}\right) = 0.
\end{equation}

\noindent Similarly, conservation of mass for the fluid is

\begin{equation} \label{gen:fluid-mass}
    \frac{\p \phi_{f}}{\p t} + \bm{\nabla}\boldsymbol{\cdot}\left(\phi_{f}\boldsymbol{v}_{f}\right) = 0,
\end{equation}

\noindent where $\boldsymbol{v}_{f}$ is the fluid velocity. Darcy's law for the fluid in the porous medium is given by

\begin{equation} \label{gen:fluid-darcy}
    \phi_{f}\left(\boldsymbol{v}_{f} - \boldsymbol{v}_{s}\right) = -\frac{k(\phi_{f})}{\mu}\left(\bm{\nabla}p_{f} - \rho_{f}\boldsymbol{g}\right),
\end{equation}

\noindent which relates the velocity of the fluid relative to the solid to the permeability of the solid, $k$, the dynamic viscosity of the fluid, $\mu$, and fluid density, $\rho_{f}$, and the body force per unit total mass, $\boldsymbol{g}$. We use the Kozeny-Carman formula~\citep{Kozeny1927,Carman1937},

\begin{equation} \label{gen:fluid-k}
    k(\phi_{f}) = k_{0}\frac{\left(1 - \phi_{f,0}\right)^{2}}{\phi_{f,0}^{3}}\frac{\phi_{f}^{3}}{\left(1 - \phi_{f}\right)^{2}},
\end{equation}

\noindent to model the permeability of the porous medium. The reference permeability $k_{0}$ corresponds to the permeability of the reference material with uniform fluid fraction $\phi_{f,0}$, and we take it to be proportional to the square of the pore size, $d$, via

\begin{equation} \label{gen:fluid-k_0}
    k_{0} = \frac{\phi_{f,0}^{3}}{\left(1 - \phi_{f,0}\right)^{2}}\frac{d^{2}}{180}.
\end{equation}

Conservation of linear momentum is given by

\begin{equation} \label{gen:mech-navier}
    \bm{\nabla}\boldsymbol{\cdot}\bm{\upsigma} = -\rho\boldsymbol{g},
\end{equation}

\noindent where $\rho$, the phase-averaged density, is defined in terms of the fluid density, $\rho_{f}$, and solid density, $\rho_{s}$, via

\begin{equation} \label{gen:mech-rho}
    \rho = \phi_{f}\rho_{f} + (1 - \phi_{f})\rho_{s}.
\end{equation}

\noindent In (\ref{gen:mech-navier}), we neglect inertia, since weakening of the solid skeleton is slow relative to elastic wave propagation. For a poroelastic material, the total stress can be conveniently separated into the Terzaghi effective stress, $\bm{\upsigma}'$, and the fluid pressure, $p_{f}$, through

\begin{equation} \label{gen:const-sigma}
    \bm{\upsigma} = \bm{\upsigma}' - p_{f}\mathsfbi{I}.
\end{equation}

\noindent The Terzaghi effective stress represents the contribution to the total stress arising from elastic deformations of the solid. For the constitutive relation for the solid, we choose the following neo-Hookean model (see e.g.~\citet{Hennessy2022}):

\begin{equation} \label{gen:const-const}
    \bm{\upsigma}' = \frac{\upnu E}{(1 + \upnu)(1 - 2\upnu)}(J - 1)\mathsfbi{I} + \frac{E}{2(1 + \upnu)J}\left(\mathsfbi{F}\mathsfbi{F}^{\mathrm{T}} - \mathsfbi{I}\right).
\end{equation}

\noindent The variable $E$ and constant $\upnu$ are the Young's modulus and Poisson's ratio, respectively. We assume that the Young's modulus decays over time in response to the solute interacting with the skeleton but that the Poisson's ratio remains fixed. 

Equations (\ref{gen:solid-phi_s})--(\ref{gen:const-const}) represent the governing equations for a poroelastic material undergoing large deformations. We now introduce a chemical species dissolved in the interstitial fluid that reduces the stiffness of the solid matrix, which we capture via the Young's modulus. We assume that the solute is transported through the poroelastic medium by advection and diffusion, and we denote its concentration per unit fluid volume as $c$. We further assume that there is an adequate supply of solute to the system. The concentration satisfies an advection-diffusion equation, as described by~\citet{Fiori2025}, given by

\begin{equation} \label{gen:weak-c}
    \frac{\p }{\p t}(\phi_{f}c) = \bm{\nabla}\boldsymbol{\cdot}\left[\phi_{f}\D{m}\bm{\nabla}c - \phi_{f}c\boldsymbol{v}_{f}\right],
\end{equation}

\noindent where $\D{m}$ is the coefficient of molecular diffusion of the solute in pure solvent, which we take to be constant. In this model, we neglect the role of dispersion. We also neglect the effects of solute consumption through this process, which would introduce an additional term of the form $-\kappa\phi_fc$ to the right-hand side of (\ref{gen:weak-c}), where $\kappa$ is the first-order rate of reaction. This choice is motivated by two distinct model assumptions. Firstly, if the rate of reaction is much smaller than the rate of solute advection or diffusion, then any consumed solute will be rapidly replenished. Secondly, if the solute is a catalyst, such as an enzyme, it is not consumed as it weakens the solid skeleton.

We suppose that the solute interacts with the solid skeleton and causes it to weaken, and that the overall rate of weakening is proportional to the chemical concentration. Using a concentration-dependent rate of weakening is motivated by~\citet{AbiAkl2019} and~\citet{Vernerey2012}. We further suppose that the Young's modulus of the material is bounded below via a minimal stiffness, $\Emin$, and model the evolution of the Young's modulus, $E$, as follows:

\begin{equation} \label{gen:weak-E}
    \frac{\p E}{\p t} + \boldsymbol{v}_{s}\cdot\bm{\nabla}E = -\beta_{E}c\left(E - \Emin\right),
\end{equation}

\noindent where $\beta_{E}$ is the decay rate of the Young's modulus per unit of concentration of solute. 

\section{Modelling flow-driven uniaxial compression}\label{sec:channel-model}

We now present the one-dimensional version of the three-dimensional system in \S\ref{sec:3D-model}, together with the initial and boundary conditions for a poroelastic material in a channel. The one-dimensional geometry we consider facilitates examination of the fundamental physics of the model. Furthermore, one-dimensional systems are quick to simulate numerically and reveal the important underlying mechanisms of the system. We condense and nondimensionalise the model, whereby the simplified formulation retains only the porosity, Young's modulus and solute concentration. We extract the fundamental timescale ratios of the nondimensional model; these key parameters are varied in the remainder of the paper to capture various physical applications.

\begin{figure*}
 \centering
 \includegraphics[height=7.5cm]{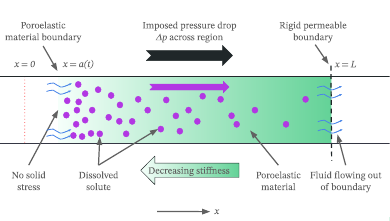}
 \caption{Diagram of a weakening poroelastic material of initial length $L$. The poroelastic material (represented by a green rectangle) is fixed to a rigid permeable wall on the right at $x = L$ and is free to deform on the left at $x = a$. A solute (represented by purple circles) is carried by the interstitial fluid of a poroelastic medium, which is driven by a drop in pressure across the domain. The solute weakens the material, where lighter shades of green represent reduced stiffness.}
 \label{fig:poroelastic-material}
\end{figure*}

Specifically, we analyse a poroelastic material in a channel of initial length $L$ undergoing uniaxial deformation, subject to a fixed drop in fluid pressure across the material, $\Delta p$ (see figure~\ref{fig:poroelastic-material}). We denote the single spatial Eulerian coordinate by $x$ with the unit vector $\boldsymbol{e}_x$ in the direction of increasing $x$. The left boundary of the material, at $x = a(t)$, is free to move, and we assume continuity of total stress and fluid stress at this boundary. The right boundary, at $x = L$, is attached to a rigid, permeable wall. The pressure drop $\Delta p$ drives the flow of the interstitial fluid through the pore space from left to right, which in turn causes deformation of the porous structure. We suppose that to the left of the poroelastic domain, a well-mixed fluid at pressure $\Delta p$ at $x = a(t)$ carries a solute of constant concentration $c^{*}$ into the poroelastic medium. The dissolved solute, represented by purple circles in figure~\ref{fig:poroelastic-material}, is transported via the interstitial fluid from the left boundary and advects and diffuses through the domain. The higher the local concentration of solute, the greater the rate at which the material weakens; we represent this with a lightening colour gradient from right to left in the diagram.

For this one-dimensional material, we assume the following: the solid skeleton deforms uniaxially, all dependent variables are functions of $x$ and $t$, and there is zero applied body force. Therefore, the solid displacement, solid velocity and fluid velocity take the forms

\begin{equation} \label{1D:vec-to-scal}
    \boldsymbol{u}_{s}(\boldsymbol{x}, t) = u_{s}(x, t)\boldsymbol{e}_{x}, \qquad \boldsymbol{v}_{s}(\boldsymbol{x}, t) = v_{s}(x, t)\boldsymbol{e}_{x}, \qquad \boldsymbol{v}_{f}(\boldsymbol{x}, t) = v_{f}(x, t)\boldsymbol{e}_{x}.
\end{equation}

\noindent We also denote the axial total stress and Terzaghi stress as $\sigma_{xx}$ and $\sigma_{xx}'$, respectively.

\subsection{Boundary and initial conditions}

Since the left boundary is moving and the right boundary is fixed (see figure~\ref{fig:poroelastic-material}), the solid displacement satisfies the following boundary conditions:

\vspace{-0.5cm}
\begin{subequations} \label{1D:BCs-u_s}
\begin{align}
\begin{split} \label{1D:BCs-u_s-a}
    u_{s} = a \qquad \mathrm{on} \quad x &= a(t),
\end{split}\\
\begin{split} \label{1D:BCs-u_s-L}
    u_{s} = 0 \qquad \mathrm{on} \quad x &= L.
\end{split}
\end{align}
\end{subequations}

\noindent Without loss of generality, we define $p_{f}$ relative to the fluid pressure at $x = L$. As described above, we apply continuity of total stress at $x =a$. The dynamic conditions for our model are thus

\vspace{-0.5cm}
\begin{subequations} \label{1D:BCs-p_f}
\begin{align}
\begin{split} \label{1D:BCs-p_f-a}
    p_{f} = \Delta p, \qquad \sigma_{xx} = -\Delta p \qquad \mathrm{on} \quad x &= a(t),
\end{split}\\
\begin{split} \label{1D:BCs-p_f-L}
    p_{f} = 0 \qquad \mathrm{on} \quad x &= L.
\end{split}
\end{align}
\end{subequations}

\noindent The conditions on the solute concentration are 

\vspace{-0.5cm}
\begin{subequations} \label{1D:BCs-c}
\begin{align}
\begin{split} \label{1D:BCs-c-a}
    \phi_{f}\left(\D{m}\frac{\p c}{\p x} - \left(v_{f} - \dot{a}\right)\left(c - c^{*}\right)\right) = 0 \qquad \mathrm{on} \quad x &= a(t),
\end{split}\\
\begin{split} \label{1D:BCs-c-L}
    \frac{\p c}{\p x} = 0 \qquad \mathrm{on} \quad x &= L.
\end{split}
\end{align}
\end{subequations}

\noindent Equation (\ref{1D:BCs-c-a}) is conservation of total solute flux at the left boundary, and (\ref{1D:BCs-c-L}) is no diffusive flux of solute at the right boundary. See appendix~\ref{sec:appendix-c-ibcs} for a detailed derivation of the boundary conditions on $c$.

We now use the model equations to convert conditions on $u_s$, $p_f$ and $\sigma_{xx}$ into conditions on $v_s$ and $\sigma_{xx}'$; this will be useful in reformulating the model below. We insert conditions (\ref{1D:BCs-u_s}) into (\ref{gen:solid-v_s}) to get that

\vspace{-0.5cm}
\begin{subequations} \label{1D:BCs-kin}
\begin{align}
\begin{split} \label{1D:BCs-kin-a}
    v_{s} = \dot{a} \qquad \mathrm{on} \quad x &= a(t),
\end{split}\\
\begin{split} \label{1D:BCs-kin-L}
    v_{s} = 0 \qquad \mathrm{on} \quad x &= L.
\end{split}
\end{align}
\end{subequations}

\noindent It is useful to integrate the $x$ component of the equation for mechanical equilibrium (\ref{gen:mech-navier}) in the absence of the body force and use the three boundary conditions (\ref{1D:BCs-dyn}) along with the relation (\ref{gen:const-sigma}) to determine the following conditions for the Terzaghi stress:

\vspace{-0.5cm}
\begin{subequations} \label{1D:BCs-dyn}
\begin{align}
\begin{split} \label{1D:BCs-dyn-a}
    \sigma_{xx}' = 0 \qquad \mathrm{on} \quad x &= a(t),
\end{split}\\
\begin{split} \label{1D:BCs-dyn-L}
    \sigma_{xx}' = -\Delta p \qquad \mathrm{on} \quad x &= L.
\end{split}
\end{align}
\end{subequations}

\noindent We use conditions (\ref{1D:BCs-kin}) and (\ref{1D:BCs-dyn}) in place of (\ref{1D:BCs-u_s}) and (\ref{1D:BCs-p_f}), respectively, in the analysis that follows.

The initial conditions for the left boundary position, porosity, Young's modulus, and solute concentration are

\begin{equation} \label{1D:ICs}
    a = 0, \qquad \phi_{f} = \phi_{f,0}, \qquad E = E_{0}, \qquad c = 0 \qquad \mathrm{at} \quad t = 0,
\end{equation}

\noindent where $E_0$ is the initial Young's modulus.

\subsection{Model reformulation and nondimensionalisation} \label{sec:nondimensional}

We now briefly present the nondimesional system of equations. See appendix~\ref{sec:appendix-deriving-condensed-model} for the full derivation. Before nondimensionalising, we introduce the phase-averaged velocity, $v$, as

\begin{equation} \label{1D:v}
    v = \phi_{f}v_{f} + (1 - \phi_{f})v_{s}.
\end{equation}

\noindent We nondimensionalise the variables as follows, using tildes to denote dimensionless quantities:

\vspace{-0.5cm}
\begin{equation} \label{nd:scalings}
\begin{split}
    x = L\tilde{x}, \qquad a = L\tilde{a}, \qquad t &= t_{\mathrm{weak}}\tilde{t}, \qquad k = k_{0}\tilde{k}, \qquad \sigma_{xx}' = E_{0}\tilde{\sigma}_{xx}',\\ \qquad (v, v_{f}, v_{s}) = v^{*}(\tilde{v}, \tilde{v}_{f}, \tilde{v}_{s}), \qquad E &= E_{0}\tilde{E}, \qquad \Emin = E_0\tilde{E}_{\mathrm{min}}  \qquad c = c^{*}\tilde{c}, 
\end{split}
\end{equation}

\noindent where

\begin{equation} \label{nd:vstar}
    v^{*} := \frac{k_{0}\Delta p}{\mu L}
\end{equation}

\noindent is the characteristic velocity obtained from Darcy's law (\ref{gen:fluid-darcy}) and

\begin{equation} \label{nd:t_weak}
    t_{\mathrm{weak}} := \frac{1}{\beta_E c^{*}}
\end{equation}

\noindent is the timescale of weakening. In (\ref{nd:scalings}), we choose to nondimensionalise time with $t_{\mathrm{weak}}$, since we are primarily interested in the dynamics of the material as it weakens. For ease of notation, we drop tildes for the remainder of this work when referring to nondimensional variables. 

The nondimensional system of equations is

\vspace{-0.5cm}
\begin{subequations} \label{nd:system-condensed}
\begin{align}
\begin{split} \label{nd:system-condensed-phi}
    \epsilon\frac{\mathrm{D}\phi_{f}}{\mathrm{D}{t}} &= \frac{\p}{\p x}\left[(1 - \phi_{f})k(\phi_{f})\frac{\p}{\p x}\left(Eg(\phi_{f})\right)\right],
\end{split}\\
\begin{split} \label{nd:system-condensed-v}
    \frac{\p v}{\p x} &= 0,
\end{split}\\
\begin{split} \label{nd:system-condensed-E}
    \frac{\mathrm{D}^{s}E}{\mathrm{D}t} &= -c\left(E - \Emin\right),
\end{split}\\
\begin{split} \label{nd:system-condensed-c}
    \Pen\, \epsilon\, \gamma^{-1}\phi_{f}\frac{\mathrm{D}^{f}c}{\mathrm{D}t} &= \frac{\p}{\p x}\left(\phi_{f}\frac{\p c}{\p x}\right),
\end{split}
\end{align}
\end{subequations}

\noindent where we define

\begin{equation} \label{nd:g}
    g\left(\phi_{f}\right) := \frac{1}{2(1+\upnu)(1-2\upnu)}\left[\frac{1-\phi_{f,0}}{1-\phi_{f}}-2\upnu-(1-2\upnu)\frac{1-\phi_{f}}{1-\phi_{f,0}}\right],
\end{equation}

\noindent and

\vspace{-0.5cm}
\begin{subequations} \label{nd:mater-deriv}
\begin{align}
\begin{split} \label{nd:D_Dt}
    \frac{\mathrm{D}}{\mathrm{D}t} &:= \frac{\p}{\p t} + \epsilon^{-1}\gamma v\frac{\p}{\p x},
\end{split}\\
\begin{split} \label{nd:Ds_Dt}
    \frac{\mathrm{D}^{s}}{\mathrm{D}t} &:= \frac{\p}{\p t} + \epsilon^{-1}\gamma v_{s}\frac{\p}{\p x},
\end{split}\\
\begin{split} \label{nd:Df_Dt}
    \frac{\mathrm{D}^{f}}{\mathrm{D}t} &:= \frac{\p}{\p t} + \epsilon^{-1}\gamma v_{f}\frac{\p}{\p x},
\end{split}
\end{align}
\end{subequations}

\noindent are the total material derivative, solid material derivative and fluid material derivative, respectively. The function $g$ can be interpreted as a nonlinear elastic strain, which can be seen from the nondimensional neo-Hookean stress-strain relation

\begin{equation} \label{nd:sigma_xx'-phi}
    \sigma'_{xx} = Eg(\phi_{f}).
\end{equation}

\noindent The solid and fluid velocities appearing in (\ref{nd:mater-deriv}) can be written explicitly in terms of $\phi_{f}$, $E$ and $v$; see appendix~\ref{sec:appendix-deriving-condensed-model}. The nondimensional parameters $\epsilon$, $\Pen$ and $\gamma$ defined in (\ref{nd:system-condensed}) are

\begin{equation} \label{nd:epsilon-Pe-gamma}
    \epsilon:=\frac{t_{\mathrm{pe}}}{t_{\mathrm{weak}}}, \qquad \Pen:=\frac{t_{\mathrm{diff}}}{t_{\mathrm{adv}}},\qquad \gamma:=\frac{t_{\mathrm{pe}}}{t_{\mathrm{adv}}},
\end{equation}

\noindent where

\begin{equation} \label{nd:timescales}
    t_{\mathrm{pe}} := \frac{\mu L^{2}}{k_{0}E_{0}}, \qquad t_{\mathrm{adv}} := \frac{L}{v^{*}} = \frac{\mu L^{2}}{k_{0}\Delta p}, \qquad t_{\mathrm{diff}} := \frac{L^{2}}{\D{m}},
\end{equation}

\noindent represent the poroelastic relaxation timescale, $t_{\mathrm{pe}}$, the advection timescale, $t_{\mathrm{adv}}$, and the solute diffusion timescale, $t_{\mathrm{diff}}$, respectively. The poroelastic relaxation timescale captures the time it takes for the material to equilibrate in response to an external stress or deformation. Together with the weakening timescale, $t_{\mathrm{weak}}$, the interplay between these four timescales dictates the dynamics of the system, and they vary greatly between different physical applications.

We use (\ref{nd:sigma_xx'-phi}) to convert the conditions on the Terzaghi stress (\ref{1D:BCs-dyn}) into conditions on the porosity and Young's modulus to yield

\vspace{-0.5cm}
\begin{subequations} \label{nd:BCs-phi}
\begin{align}
\begin{split} \label{nd:BCs-phi-a}
    \phi_{f} = \phi_{f,0} \qquad &\mathrm{on} \quad x = a,
\end{split}\\
\begin{split} \label{nd:BCs-phi-1}
    Eg(\phi_{f}) = -\gamma \qquad &\mathrm{on} \quad x = 1.
\end{split}
\end{align}
\end{subequations}

\noindent Equation (\ref{nd:BCs-phi-a}) is a consequence of the fact that $g(\phi_{f})=0$ implies $\phi_f = \phi_{f,0}$. We note from (\ref{nd:epsilon-Pe-gamma})--(\ref{nd:timescales}) that $\gamma$ can also be written as $\Delta p / E_{0}$, the ratio of the pressure drop to the initial material stiffness. The dimensionless boundary conditions on the solute concentration (\ref{1D:BCs-c}) are

\vspace{-0.5cm}
\begin{subequations} \label{nd:BCs-c}
\begin{align}
\begin{split} \label{nd:BCs-c-a}
    \Pen^{-1}\frac{\p c}{\p x} - \left(v_{f} - \dot{a}\right)\left(c - 1\right) = 0 \qquad \mathrm{on} \quad x &= a,
\end{split}\\
\begin{split} \label{nd:BCs-c-1}
    \frac{\p c}{\p x} = 0 \qquad \mathrm{on} \quad x &= 1.
\end{split}
\end{align}
\end{subequations}

The dimensionless initial conditions (see (\ref{1D:ICs})) are

\begin{equation} \label{nd:ICs}
    a = 0, \qquad \phi_{f} = \phi_{f,0}, \qquad E = 1, \qquad c = 0 \qquad \mathrm{at} \quad t = 0.
\end{equation}

\noindent In appendix~\ref{sec:appendix-deriving-condensed-model}, we use the kinematic conditions (\ref{1D:BCs-kin}) to derive equations to determine the phase-averaged velocity and the position of the free boundary given by

\vspace{-0.5cm}
\begin{subequations} \label{nd:constraints}
\begin{align}
\begin{split} \label{nd:constraint-v}
    \gamma v &= -k(\phi_{f})\frac{\p}{\p x}\left(Eg(\phi_{f})\right)\Bigg|_{x=1},
\end{split}\\
\begin{split} \label{nd:constraint-a}
    a &= \phi_{f,0} - \int_{a}^{1}\phi_{f}\ \mathrm{d}x.
\end{split}
\end{align}
\end{subequations}

\noindent We solve equations (\ref{nd:constraint-v})--(\ref{nd:constraint-a}) in conjunction with the system (\ref{nd:system-condensed}) and boundary conditions (\ref{nd:BCs-phi})--(\ref{nd:BCs-c}).

\subsection{Parameter values for applications} \label{sec:params}

We use our mathematical model to simulate the dynamical behaviour of two types of poroelastic materials undergoing weakening in the presence of a chemical species. The two examples involve (I) a porous polymer degrading in simulated body fluid (SBF) and (II) a polymer hydrogel that degrades in the presence of an enzyme. We select these two applications as they have different underlying timescales. The baseline parameters for each example are listed in table~\ref{table:parameter-vals}.  

\begin{table}
  \begin{center}
\def~{\hphantom{0}}
  \begin{tabular*}{\textwidth}{@{\extracolsep{\fill}}lllll}
    \hline
    Symbol & Description & Unit & (I) Porous PLA/PCL polymer & (II) Dex-MA hydrogel \\
    \hline
    $L$ & Lengthscale & m & $4.5\times 10^{-2}$ & $2.0\times 10^{-3}$\\
    $\phi_{f,0}$ & Initial porosity & --- & 0.50 & 0.50\\
    $k_{0}$ & Permeability scale & m\textsuperscript{2} & $5.6\times 10^{-15}$ & $3.1\times 10^{-19}$\\
    $\D{m}$ & Solute diffusion coefficient & m\textsuperscript{2}s\textsuperscript{-1} & $1.0\times 10^{-9}$ & $1.6\times 10^{-10}$\\
    $\Delta p$ & Pressure drop & Pa & $1.4\times 10^{4}$ & $7.9\times 10^{4}$\\
    $E_{0}$ & Initial Young's modulus & Pa & $1.4\times 10^{5}$ & $1.6\times 10^{5}$\\
    $\beta_{E}$ & Weakening parameter & Lg\textsuperscript{-1}s\textsuperscript{-1} & $3.3\times 10^{-7}$ & $1.2\times 10^{-2}$\\
    $c^{*}$ & Solute concentration scale & gL\textsuperscript{-1} & $3.3$ & $4.1\times 10^{-4}$\\
    $t_{\mathrm{weak}}$ & Weakening timescale & s & $9.1\times 10^{5}$ & $2.1\times 10^{5}$\\
    $t_{\mathrm{pe}}$ & Poroelastic timescale & s &  $2.6\times 10^{3}$ & $8.1\times 10^{4}$\\
    $t_{\mathrm{adv}}$ & advection timescale & s & $1.3\times 10^{4}$ & $1.6\times 10^{5}$\\
    $t_{\mathrm{diff}}$ & Solute diffusion timescale & s & $2.0\times 10^{6}$ & $2.5\times 10^{4}$\\
    \hline
  \end{tabular*}
  \caption{\ Baseline parameter values for weakening poroelastic materials. Parameters for example I primarily come from~\citet{LeBlon2013} and~\citet{Kokubo1990}. in example II, see~\citet{Meyvis1999,Stenekes2000,Hennink1997,Sun2023}. See (\ref{nd:scalings}) for definitions of the timescales.}
  \label{table:parameter-vals}
  \end{center}
\end{table}

In all simulations, we set the Poisson's ratio to be $0.3$~\citep{Javanmardi2021} and the fluid viscosity to be $10^{-3}\mathrm{\ Pa \cdot s}$ --- the viscosity of water~\citep{Berstad1988}. For the permeabilities, we used (\ref{gen:fluid-k_0}) to convert from the pore size, $d$, to the permeability scale, $k_{0}$. 

The example I parameter set is based on the experiments of~\citet{LeBlon2013}, who used a salt-leaching technique to create macropores in various polymers, including a poly(lactic acid)/poly($\mathrm{\epsilon}$-caprolactone) (PLA/PCL) blend, which we consider here. They submerged these porous polymers in SBF (a substance containing various ionic compounds) over a period of weeks to investigate how the materials weakened. To facilitate computing numerical solutions, whilst retaining a macroporous system, we decrease the pore size $d$ from the quoted $200\ \mathrm{\mu m}$~\citep{LeBlon2013} to $1\ \mathrm{\mu m}$. Larger pore sizes, and hence larger permeabilities, decrease the value of $\epsilon$ (see (\ref{nd:epsilon-Pe-gamma})--(\ref{nd:timescales})), meaning that the system needs to be simulated at smaller times to capture the full behaviour of the porosity. However, simulating at smaller times leads to boundary layers in $c$, as the diffusion lengthscale becomes much smaller than the mesh size. As seen in table~\ref{table:parameter-vals}, the poroelastic and advection timescales are the smallest (on the order of hours). Conversely, the weakening and solute diffusion timescales are on the order of days; as a result, the system is in a regime where the material weakens slowly, and the evolution of the solute is advection-dominated. Given these timescales, the nondimensional timescale ratios separate, such that $\epsilon\ll\gamma\ll\Pen$ (see table~\ref{table:dless-parameter-vals}).

\begin{table}  
    \begin{center}
\def~{\hphantom{0}}
    \begin{tabular}{@{\extracolsep{\fill}}llll}
    \hline
    Formula & Unit & Example I & Example II \\
    \hline
    $\phi_{f,0}$ & --- & 0.50, 0.60 & 0.50\\
    $\Delta p$ & Pa & $1.4\times 10^{4}$, $2.8\times 10^{4}$ & $7.9\times 10^{4}$\\
    $c^{*}$ & gL\textsuperscript{-1} & 3.3 & $4.1\times 10^{-5}$, $4.1\times 10^{-4}$, $4.1\times 10^{-3}$, $4.1\times 10^{-2}$\\
    $\Emin$ & --- & 0.20 & 0.70\\
    $\gamma$ & --- & 0.10, 0.20 & 0.50\\
    $\Pen$ & --- & 78, $1.6\times 10^{2}$ & 0.16\\
    $t_{\mathrm{pe}}/t_{\mathrm{weak}}=\epsilon$ & --- & $2.8\times 10^{-3}$ & 0.038, 0.38, 3.8, 38\\
    $t_{\mathrm{adv}} / t_{\mathrm{weak}}=\epsilon\,\gamma^{-1}$ & --- & $2.8\times 10^{-2}$, $1.4\times 10^{-2}$ & 0.077, 0.77, 7.7, 77\\
    $t_{\mathrm{diff}} / t_{\mathrm{weak}}=\Pen\,\epsilon\,\gamma^{-1}$ & --- & 2.2 & 0.012, 0.12, 1.2, 12\\
    \hline
    \end{tabular}
\caption{
{\ Key parameter values varied across examples I -- II and their effects on the dimensionless parameters of the system.}}
\label{table:dless-parameter-vals}
    \end{center}
\end{table}

The example II parameter set is based on the work of~\citet{Meyvis1999}, who studied the rheology and degradation of Dex-MA hydrogels in the presence of the enzyme dextranase. They varied the concentration of dextranase across experiments and found in some cases that doubling the concentration doubled the overall rate of weakening, which aligns with our modelling choice for the weakening law (\ref{gen:weak-E}). In our simulations, we suppose that water carries the enzyme through the hydrogel. All baseline timescales of this simulation are of a similar duration (see table~\ref{table:parameter-vals}), meaning that the coupling between poroelastic deformation, solute transport and weakening is stronger. Consequently, the baseline values of $\epsilon$, $\Pen$ and $\gamma$ are also all of a similar size. The main solute transport mechanism is diffusion in this simulation, since $t_{\mathrm{diff}}$ is shorter than $t_{\mathrm{adv}}$, though advection still plays a role.

\subsection{Finite element simulations}

We use the finite element method to simulate the nondimensional system described in \S\ref{sec:nondimensional} with baseline parameter values for examples I and II listed in table~\ref{table:parameter-vals}. The finite element method is implemented using FEniCS~\citep{Logg2012,Alnaes2015}, an open-source Python package. See appendix~\ref{sec:appendix-fenics} for details of the numerical set-up and appendix~\ref{sec:appendix-stss} for a small-time similarity solution, which we use to initialise the numerical simulations. Specifically, rather than imposing the initial conditions in (\ref{nd:ICs}), we evaluate the similarity solution at a small time, which enhances convergence.

We plot all dependent variables against the Lagrangian coordinate $X = x - u_{s}$. In addition, we calculate the time-varying spatially averaged quantities $\overline{\phi_{f}}$, $\overline{E}$ and $\overline{c}$ for the porosity, Young's modulus and solute concentration, respectively, using the following formula:

\begin{equation} \label{nd:spatial-average}
    \overline{f}(t):=\frac{1}{1-a}\int_{x=a}^{1}f(x, t)\mathrm{d} x,
\end{equation}

\noindent where $f = \phi_f,E,c$.

\section{Steady-state analysis}\label{sec:steady-state}

We now analyse the steady state of the nondimensional system (\ref{nd:system-condensed}), subject to boundary conditions (\ref{nd:BCs-phi})--(\ref{nd:BCs-c}) and equations (\ref{nd:constraints}). We identify critical parameter groups at which the interaction between fluid-driven compression and weakening leads to pore closure at the rigid boundary of the domain. Firstly, the steady-state solution requires $v_{s} \equiv 0$ (by (\ref{gen:solid-v_s})) and $c \equiv 1$, the latter of which implies that $E \equiv \Emin$ from (\ref{nd:system-condensed-E}). Consequently, using (\ref{nd:system-condensed-phi}), the steady-state equation for the porosity $\phi_{f}$ is

\begin{equation} \label{ss:system-phi}
    \gamma v\frac{\mathrm{d}\phi_{f}}{\mathrm{d}x} = \Emin\frac{\mathrm{d}}{\mathrm{d}x}\left[(1 - \phi_{f})k(\phi_{f})\frac{\mathrm{d}}{\mathrm{d} x}\left(g(\phi_{f})\right)\right],
\end{equation}

\noindent where $v$ is a constant to be determined (see (\ref{nd:system-condensed-v})). The boundary conditions (\ref{nd:BCs-phi}) and equations (\ref{nd:constraints}) remain unchanged and close this system. We restate them here for convenience as:

\vspace{-0.5cm}
\begin{subequations} \label{ss:BCs-phi}
\begin{align}
\begin{split} \label{ss:BCs-phi-a}
    \phi_{f} = \phi_{f,0} \qquad &\mathrm{on} \quad x = a,
\end{split}\\
\begin{split} \label{ss:BCs-phi-1}
    Eg(\phi_{f}) = -\gamma \qquad &\mathrm{on} \quad x = 1,
\end{split}
\end{align}
\end{subequations}

\noindent and

\vspace{-0.5cm}
\begin{subequations} \label{ss:constraints}
\begin{align}
\begin{split} \label{ss:constraint-v}
    \gamma v &= -k(\phi_{f})\frac{\mathrm{d}}{\mathrm{d}x}\left(Eg(\phi_{f})\right)\Bigg|_{x=1},
\end{split}\\
\begin{split} \label{ss:constraint-a}
    a &= \phi_{f,0} - \int_{a}^{1}\phi_{f}\ \mathrm{d}x.
\end{split}
\end{align}
\end{subequations}

\noindent We note that $a$ and $v$ are unknown scalars which must be determined as part of the solution. We integrate (\ref{ss:system-phi}) once, using (\ref{ss:constraint-v}), to get

\begin{equation} \label{ss:vs-0}
    \gamma v = -\Emin k(\phi_{f})g'(\phi_f)\frac{\mathrm{d}\phi_f}{\mathrm{d}x}.
\end{equation}

\noindent Since $v$, $k(\phi_f)$ and $g'(\phi_f)$ are all positive, we see from (\ref{ss:vs-0}) that the porosity is smallest at the right of the domain. In addition, we require that $\phi_{f} > 0$, meaning that $g(\phi_f) > g(0)$. Given this constraint, we determine that a necessary condition for the existence of a steady state, using (\ref{nd:g}) and (\ref{ss:BCs-phi-1}), is that

\begin{equation} \label{ss:alpha}
    \alphamin := \frac{\Emin}{\gamma}\frac{\phi_{f,0}(2(1-\upnu) - \phi_{f,0})}{2(1-\phi_{f,0})(1+\upnu)(1-2\upnu)} > 1.
\end{equation}

\noindent If condition (\ref{ss:alpha}) does not hold, then the pores close on the right edge of the domain before a steady state is reached, and the model breaks down since $\phi_f = 0$ and would become negative in the steady state. We recall that $\gamma = \Delta p / E_0$, and so modifying any of the initial porosity, minimum Young's modulus, Poisson's ratio or pressure drop can result in pore closure. In \S\ref{sec:introduction}, we hypothesised that weakening the material decreases the maximal pressure drop that can be sustained across the material without inducing pore closure. We see that in reducing $\Emin$, the condition (\ref{ss:alpha}) permits a smaller range of $\gamma$ and hence $\Delta p$ values. Therefore, weakening causes the range of pressure drops that lead to pore closure to increase. 

Assuming that (\ref{ss:alpha}) holds and no pore closure occurs, we integrate (\ref{ss:vs-0}) and use the boundary condition (\ref{ss:BCs-phi-1}) to get the following implicit expression for the steady-state porosity:

\begin{equation} \label{ss:phi-f}
    \frac{\gamma v}{\Emin}\frac{\phi_{f,0}^{3}}{1 - \phi_{f,0}}(1 - x) = G(\phi_{f}; \phi_{f,0}, \upnu) - G(\phi_{f,r}; \phi_{f,0}, \upnu),
\end{equation}

\noindent where $G$ is given by

\begin{align} \label{ss:F}
\begin{aligned}
    &2(1 + \upnu)(1 - 2\upnu)G(\phi_{f}; \phi_{f,0}, \upnu) \\
    &= (1 - \phi_{f,0})^{2}\left[\frac{1}{3(1 - \phi_{f})^{3}}-\frac{3}{2(1 - \phi_{f})^{2}}+\frac{3}{1 - \phi_{f}}+\log(1 - \phi_{f})\right] \\
    &+ (1 - 2\upnu)\left[\frac{1}{1 - \phi_{f}} + 3\log(1 - \phi_{f}) - 3(1 - \phi_{f}) + \frac{1}{2}(1 - \phi_{f})^{2}\right],
\end{aligned}
\end{align}

\noindent and $\phi_{f,r}$ is the steady-state porosity at the right boundary. To find the steady-state values of $v$ and $a$, we first substitute $\mathrm{d}x$ in (\ref{ss:constraint-a}) with $x'(\phi_{f})\mathrm{d}\phi_{f}$ and use (\ref{ss:phi-f}) to obtain

\begin{equation} \label{ss:a-integral}
    a = \phi_{f,0} - \frac{(1 - \phi_{f,0})\Emin}{\phi_{f,0}^{3}\gamma v}\int_{\phi_{f,r}}^{\phi_{f,0}}\phi_{f}\frac{\mathrm{d}G}{\mathrm{d}\phi_{f}}\mathrm{d}\phi_{f}.
\end{equation}

\noindent The integral in (\ref{ss:a-integral}) can be evaluated explicitly, but we omit this expression for brevity. Next, we insert the Dirichlet condition on $\phi_{f}$ (\ref{ss:BCs-phi-a}) into the steady-state equation (\ref{ss:phi-f}) and eliminate $a$ between the resulting expression and (\ref{ss:a-integral}) to get the following equation for $v$:

\begin{equation} \label{ss:v}
    v = \frac{\Emin}{\phi_{f,0}^{3}\gamma}\left[G(\phi_{f,0};\phi_{f,0},\upnu) - G(\phi_{f,r};\phi_{f,0},\upnu) - \int_{\phi_{f,r}}^{\phi_{f,0}}\phi_{f}\frac{\mathrm{d}G}{\mathrm{d}\phi_{f}}\mathrm{d}\phi_{f}\right].
\end{equation}

\noindent Equation (\ref{ss:v}) can be solved for $v$, and the resulting expression can be inserted into (\ref{ss:a-integral}) to find an explicit formula for $a$ (omitted for brevity). 

In an experimental setting, given knowledge of the pressure-drop, initial porosity and Poisson's ratio, if the steady-state value of $a$ were measured, equations (\ref{ss:a-integral})--(\ref{ss:v}) could be used to estimate the minimum Young's modulus of the material, $\Emin$.

\section{Systems with slow weakening} \label{sec:slow-weakening}

In this section, we consider poroelastic materials that weaken on a much longer timescale than that of poroelastic relaxation or advection, motivated by the timescale separation we see in example I. We will see that the Young's modulus and solute concentration are spatially homogeneous to leading order in this parameter regime. Furthermore, we will find the system to be quasi-steady, in the sense that the weakening equation is the only equation to retain its time derivative at leading order. We assume the following:

\begin{equation} \label{slow-weak:tE}
    \frac{t_{\mathrm{adv}}}{t_{\mathrm{weak}}} \ll 1, \qquad \epsilon=\frac{t_{\mathrm{pe}}}{t_{\mathrm{weak}}}  \ll 1.
\end{equation}

\noindent Since $\gamma = t_{\mathrm{pe}}/t_{\mathrm{adv}}$, we note from (\ref{slow-weak:tE}) that also $\epsilon \ll \gamma$. See table~\ref{table:dless-parameter-vals} for relationships between the timescales of the system and the dimensionless parameters. We further assume that $\gamma = O(1)$ as $\epsilon \rightarrow 0$, meaning that the applied pressure drop cannot be much larger than the initial Young's modulus of the material. We assume no constraints on the Péclet number. Although $\Pen \gg 1$ in example I, the mathematical analysis is intact for a more general Péclet number.

\subsection{Mathematical analysis}\label{sec:qss}

In the limit of slow weakening, $\epsilon \rightarrow 0$, an asymptotic solution describing the quasi-steady evolution of the system can be obtained. In the full nondimensional time-dependent system (\ref{nd:system-condensed}) we assume that $t = O(1)$, as we have scaled time with $t_{\mathrm{weak}}$. We seek a dominant balance in the solid mass conservation equation 

\begin{equation}\label{qss:solid-mass}
    \epsilon\gamma^{-1}\frac{\p\phi_f}{\p t} - \frac{\p}{\p x}\left((1-\phi_f)v_s\right)=0. 
\end{equation}

\noindent If $v_{s}$ were $O(1)$, then (\ref{1D:BCs-kin-L}) and (\ref{qss:solid-mass}) would imply that $v_{s}=0$ at leading order in $\epsilon$. Hence, we rescale the solid velocity as $v_{s} = \epsilon V_{s}$ in this regime.

The full nondimensional system is then

\vspace{-0.5cm}
\begin{subequations} \label{qss:system-condensed}
\begin{align}
\begin{split} \label{qss:system-condensed-phi}
    \epsilon\frac{\p\phi_{f}}{\p t} + \gamma v \frac{\p \phi_{f}}{\p x} &= \frac{\p}{\p x}\left[(1 - \phi_{f})k(\phi_{f})\frac{\p}{\p x}\left(Eg(\phi_{f})\right)\right],
\end{split}\\
\begin{split} \label{qss:system-condensed-E}
    \frac{\p E}{\p t} + \gamma V_{s}\frac{\p E}{\p x} &= -c\left(E - \Emin\right),
\end{split}\\
\begin{split} \label{qss:system-condensed-c}
    \epsilon\,\gamma^{-1}\phi_{f}\frac{\p c}{\p t} + \phi_{f}v_{f}\frac{\p c}{\p x} &=\Pen^{-1}\frac{\p}{\p x}\left(\phi_{f}\frac{\p c}{\p x}\right).
\end{split}
\end{align}
\end{subequations}

\noindent The boundary conditions, initial conditions and equations for determining $v$ and $a$ remain unchanged from (\ref{nd:BCs-phi})--(\ref{nd:constraints}). We seek leading order solutions to system (\ref{qss:system-condensed}) and expand all dependent variables $f$ as $f \sim f^{(0)} + \epsilon f^{(1)} + \cdots$ as $\epsilon \rightarrow 0$. With this notation we have that $\phi_{f}^{(0)}v_{f}^{(0)} = v^{(0)}$, which we can substitute in (\ref{qss:system-condensed-c}). The leading-order solute concentration, $c^{(0)}$, can be found in an analogous way to \S\ref{sec:steady-state}. We integrate (\ref{qss:system-condensed-c}) at leading order and use the boundary conditions (\ref{nd:BCs-c}) to deduce that $c^{(0)} \equiv 1$. We note that $c^{(0)}\equiv 1$ for any Péclet number, meaning that, with these boundary conditions, advection- and diffusion-dominated systems have the same quasi-steady behaviour.

Explicitly, the first order hyperbolic equation for the Young's modulus is, at leading order,

\begin{equation} \label{qss:E-lo}
    \frac{\p E^{(0)}}{\p t} + \gamma V_{s}^{(0)}\frac{\p E^{(0)}}{\p x} = - \left(E^{(0)} - \Emin\right),
\end{equation}

\noindent and is subject to the initial condition

\begin{equation} \label{qss:E-ic}
    E^{(0)} = 1 \qquad \mathrm{at} \quad t = 0.
\end{equation}

\noindent Using the method of characteristics, the solution to (\ref{qss:E-lo})--(\ref{qss:E-ic}) for $E^{(0)}$ (which does not require solving the characteristic equation for $x$) is

\begin{equation} \label{qss:E-soln}
    E^{(0)} = \Emin + (1 - \Emin)e^{-t},
\end{equation}

\noindent which is spatially uniform and drives the temporal variation of $\phi_{f}$ in (\ref{qss:system-condensed-phi}). The spatial uniformity of $E^{(0)}$ is directly caused by the spatial uniformity of $c^{(0)}$. The leading-order solution for $\phi_{f}$ can be obtained from (\ref{qss:system-condensed-phi}) and is

\begin{equation} \label{qss:phi-soln}
    \frac{\gamma v}{\Emin + (1 - \Emin)e^{-t}}\frac{\phi_{f,0}^{3}}{1 - \phi_{f,0}}(1 - x) = G\left(\phi_{f}^{(0)};\phi_{f,0},\upnu\right) - G\left(\phi_{f,r}^{(0)}; \phi_{f,0},\upnu\right).
\end{equation}

\noindent The porosity at the right boundary $\phi_{f,r}^{(0)}(t)$ is now time dependent and determined by the solution to the equation

\begin{equation} \label{qss:phi-fr}
    \left(\Emin + (1 - \Emin)e^{-t}\right)g\left(\phi_{f,r}^{(0)}(t)\right) = -\gamma,
\end{equation}

\noindent which, as discussed in \S\ref{sec:steady-state}, does not always admit a solution $\phi_{f,r}^{(0)}$ that lies in the interval $(0, 1)$. We find that $\phi_{f,r}^{(0)}(t)$ is positive if and only if

\begin{equation} \label{qss:alpha}
    \left[1 + \left(\Emin^{-1} - 1\right)e^{-t}\right]\alphamin > 1,
\end{equation}

\noindent where $\alphamin$ is given by (\ref{ss:alpha}). This leads to three distinct cases for determining whether the model indicates pore closure.

Case 1 occurs when $\alphamin > 1$, which implies that the condition (\ref{qss:alpha}) holds for all time. Therefore, the porosity on the right of the domain never drops below zero, and the quasi-steady solution for $\phi_f^{(0)}$ given by (\ref{qss:phi-soln}) tends to (\ref{ss:phi-f}) as $t \rightarrow \infty$. Case 2 occurs when $\Emin < \alphamin \leq 1$, meaning that (\ref{qss:alpha}) holds until $t = \tcrit$, where

\begin{equation} \label{qss:t-crit}
    \tcrit = \log\left(\frac{\Emin^{-1} - 1}{\alphamin^{-1} - 1}\right).
\end{equation}

\noindent Physically, weakening the material causes the pores to close on the right of the domain after a finite amount of time. Case 3 occurs when $\alphamin < \Emin$. Condition (\ref{qss:alpha}) is never satisfied and the boundary condition on the right of the domain (\ref{qss:phi-fr}) always results in a negative, unphysical value for the porosity. This could correspond to applying too large a pressure drop or having too small an initial porosity, for example.

\citet{MacMinn2016} considered fluid-driven large deformations of a non-weakening poroelastic material and observed that applying too large a pressure drop led to instantaneous pore closure, which corresponds to our case 3. Through case 2, we uncover the mechanism of weakening-induced finite-time pore closure; as the material weakens over time, the maximal pressure drop permitted by the material decreases dynamically, resulting in pore closure at a finite time. In \S\ref{sec:qss-param-space} we visualise and discuss the new system behaviour that is revealed by the addition of weakening into the model.

To compute $v(t)$ and $a(t)$ in the quasi-steady regime, we follow the procedure described in \S\ref{sec:steady-state} for determining the steady values of $v$ and $a$. In expressions (\ref{ss:a-integral}) and (\ref{ss:v}), we replace the functions $\phi_{f}$ and $\Emin$ with $\phi_{f}^{(0)}$ (given by (\ref{qss:phi-soln})) and $E^{(0)}$ (given by (\ref{qss:E-soln})), respectively. In appendix~\ref{sec:appendix-validation}, we cross-validate the early-time analysis of appendix~\ref{sec:appendix-stss} and the quasi-steady state analysis of this section against the numerical results in example I. We find that the early-time solution matches the numerics well in the regime dominated by poroelastic relaxation and that the quasi-steady state solution matches the numerical solution during the weakening phase.

\subsection{Investigating regions of parameter space determined by quasi-steady analysis}\label{sec:qss-param-space}

\begin{figure*}
 \centering
 \includegraphics[height=7cm]{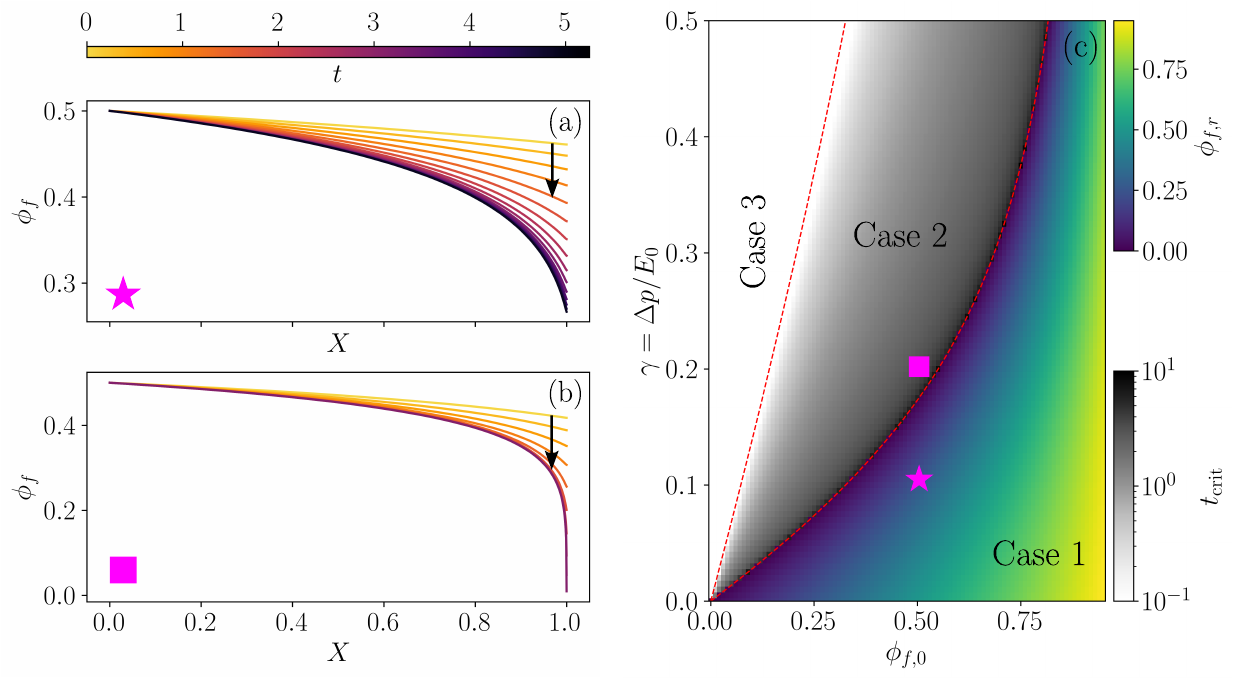}
 \caption{Analytical evolution of a slowly degrading poroelastic material. We fix $\Emin = 0.2$ and $\upnu = 0.3$. (a)--(b): Quasi-steady evolution of the porosity when $\phi_{f,0}=0.5$ and (a) $\gamma = 0.1$ and (b) $\gamma = 0.2$, corresponding to cases 1 and 2, respectively. (c): The steady-state porosity on the right boundary (case 1; (\ref{qss:phi-fr})) and time of pore closure (case 2; (\ref{qss:t-crit})) in $(\phi_{f,0}, \gamma)$ parameter space. In case 3, $\tcrit$ is equal to zero. The star and square are the locations in parameter space of the simulations in panels (a) and (b), respectively. The boundary between cases 1 and 2 is the curve $\alphamin = 1$ and the boundary between cases 2 and 3 is the curve $\alphamin = \Emin$ (see (\ref{ss:alpha})). We note that in case 2 we take the minimum and maximum values of $\tcrit$ to be $0.1$ and $10.0$, respectively, for visualisation purposes.}
 \label{fig:qss}
\end{figure*}

We now investigate cases 1--3 further by means of an example. Specifically, we set $\Emin = 0.2$, $\upnu = 0.3$, and vary the pressure drop $\Delta p$, which impacts $\gamma$, and the initial porosity $\phi_{f,0}$. We plot the quasi-steady evolution of $\phi_{f}^{(0)}$ against the Lagrangian coordinate $X = x - u_{s}^{(0)}$ for two parameter sets that illustrate cases 1 and 2 in figures~\ref{fig:qss}(a) and \ref{fig:qss}(b), respectively. The leading order displacement, $u_{s}^{(0)}$, can be found by solving (\ref{gen:solid-F})--(\ref{gen:solid-J-phi}) at leading order in $\epsilon$.

In case 1, if $\gamma$ is small enough and $\phi_{f,0}$ is large enough, the system reaches a steady state (see figure~\ref{fig:qss}(a)). In this simulation, $\alphamin = 1.73 > 1$. We plot the steady-state value of $\phi_{f,r}$ for case 1 in figure~\ref{fig:qss}(c). For each $\phi_{f,0}$, there is a corresponding $\gamma$ value for which the steady state $\phi_{f,r}$ is precisely zero, given by the red dashed curve between cases 1 and 2. This demonstrates that both a sufficiently large initial porosity and a sufficiently small pressure drop are required for the system to achieve a steady state. In addition, we see that for a fixed initial porosity (vertical lines), $\phi_{f,r}$ remains approximately constant for a large range of $\gamma$, and that only near to the boundary curve $\phi_{f,r}$ drops quickly towards zero. Therefore, a small difference in $\gamma$ can cause a large change in $\phi_{f,r}$ near the boundary curve to case 2, where the steady state is unphysical.

In case 2, weakening causes pore closure on the right of the domain at a time $\tcrit$ (see figure~\ref{fig:qss}(b)). In this simulation, $\Emin = 0.2 < \alphamin = 0.865 < 1$ and $\tcrit = 3.24$. We colour the region corresponding to case 2 according to the $\tcrit$ value in figure~\ref{fig:qss}(c) in greyscale. As the initial fluid fraction increases, the time it takes to close the pores also increases, since there is more fluid to begin with. For a fixed initial porosity, $\tcrit$ does not vary much with $\gamma$ away from the boundary between cases 1 and 2. As soon as the Young's modulus reaches its critical value, the pores close. Since $E$ decays exponentially in the quasi-steady state, the time of pore closure is similar between simulations, unless $\gamma$ is very close to its critical value.

In case 3, the pores close immediately due to the imposition of too large a pressure drop. We note that the maximal pressure drop (for a given value of $\phi_{f,0}$) is the maximal pressure drop for a system without weakening, where $\Emin = 1$.

\subsection{Numerical simulations example I: porous polymer scaffold}\label{sec:results-polymer}

In this subsection, we use the finite element method to simulate the full dynamical system described in \S\ref{sec:nondimensional}. We use parameters from~\citet{LeBlon2013} to model a porous polymer scaffold weakening in simulated body fluid (SBF), which is a system in which weakening is much slower than poroelastic relaxation. We set $\Emin = 0.2$, $\upnu = 0.3$ and $c^{*} = 3.3\ \mathrm{gL^{-1}}$, so that $\epsilon=2.8\times10^{-3}$. We take pressure drops at $10\%$ and $20\%$ of the initial Young's modulus (with $\gamma = 0.1,\ 0.2$ respectively). For each pressure drop, we simulate with an initial porosity of $0.5$ and $0.6$, which gives four distinct parameter sets in total. In these parameter sets, there is a broad separation of timescales (see table~\ref{table:dless-parameter-vals}) whereby the poroelastic timescale is smaller than the weakening and solute diffusion timescales, which are of similar orders. The advection timescale varies depending on the value of $\Delta p$ (which can be seen from (\ref{nd:timescales})).

\begin{figure*}
 \centering
 \includegraphics[height=14.2cm]{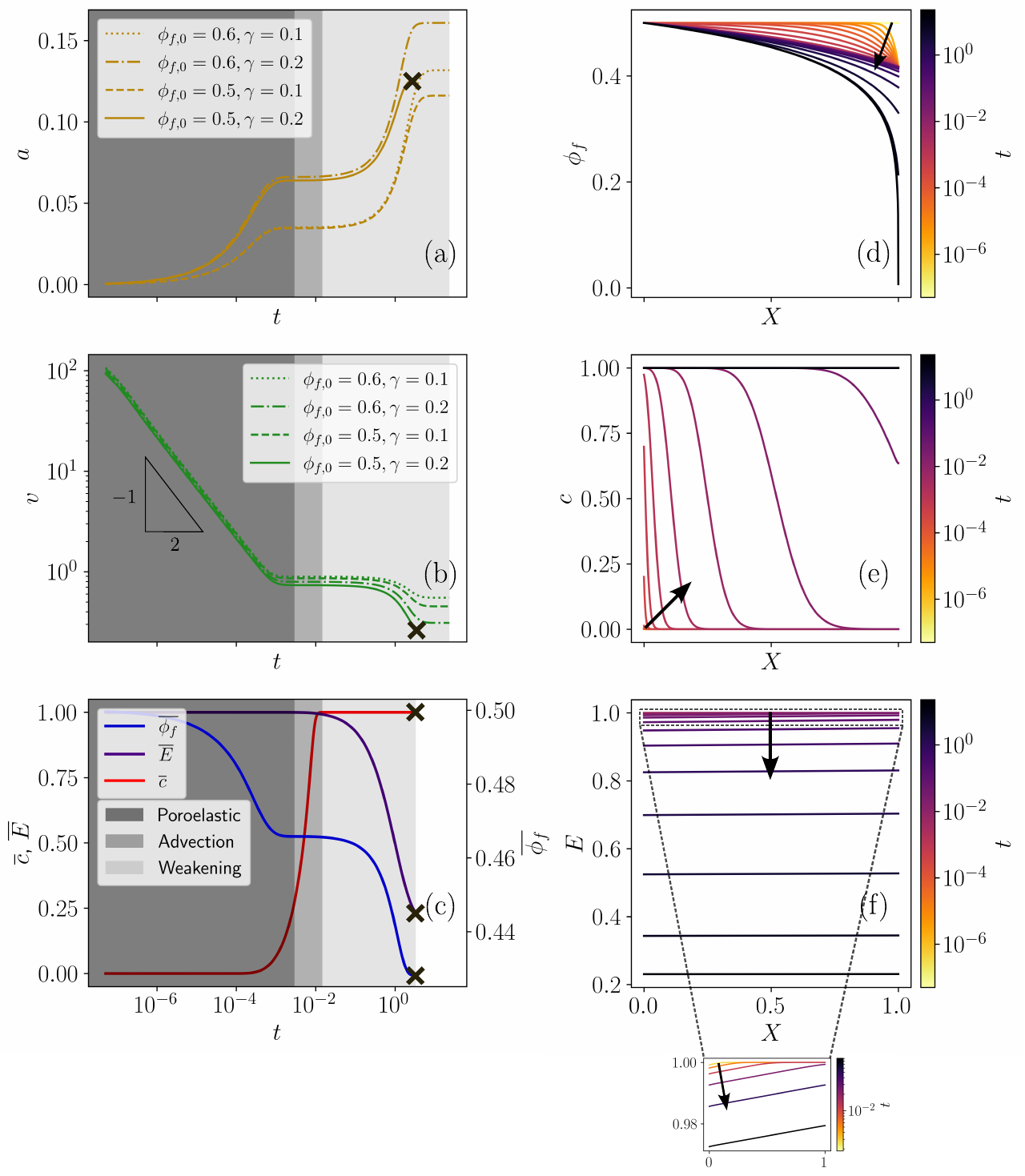}
 \caption{Numerical results for example I --- a PLA/PCL porous polymer scaffold. The evolution of (a) the left boundary, $a$ and (b) the phase-averaged velocity, $v$, over time. The dimensionless pressure drop, $\gamma$, is varied across three different simulations (denoted by dotted, dashed and solid lines). In the initial phase, the log-log plot of $v$ against $t$ is approximately linear with gradient $-\frac{1}{2}$. (c): A plot of time-varying spatially averaged quantities where $\gamma = 0.2$, $\phi_{f,0} = 0.5$. The left $y$-axis is the scale for $\overline{E}$ and $\overline{c}$, while the right $y$-axis is for $\overline{\phi_{f}}$.  In panels (a)--(c), the black crosses represent pore closure and the boundaries between shaded regions are $t_{\mathrm{pe}}/t_{\mathrm{weak}}$, $t_{\mathrm{adv}}/t_{\mathrm{weak}}$ and the simulation end time, respectively. The labels of the shaded regions refer to the dominant mechanism in the region. We shade the plot using $t_{\mathrm{adv}}/t_{\mathrm{weak}}$ when $\gamma = 0.2$ in all cases. Finally, we plot the evolution of (d) the porosity $\phi_{f}$, (e) the solute concentration $c$, and (f) the Young's modulus $E$ over time. As with panel (c), we fix $\gamma = 0.2$, $\phi_{f,0} = 0.5$ in panels (d)--(f). The arrows denote the direction of increasing time. We plot against the Lagrangian coordinate, $X$. }
 \label{fig:polymer-simulation}
\end{figure*}

Numerical simulations reveal that the left boundary of the material moves inwards in two distinct phases: an initial phase dominated by fluid-driven compression on the poroelastic timescale, followed by a second phase in which weakening dominates the deformation of the material on the longer weakening timescale (see figure~\ref{fig:polymer-simulation}(a), Movie 1). The incoming flow from the left boundary instantly pressurises the fluid across the poroelastic material. However, the requirement that the fluid pressure is zero at the right boundary leads to a boundary layer. The large pressure gradient rapidly expels fluid from the right of the material, resulting in a large phase-averaged velocity for small times (see figure~\ref{fig:polymer-simulation}(b)). During the initial poroelastic phase, $v \propto 1/\sqrt{t}$, as predicted from an analysis of the problem in the limit $t \rightarrow 0$ (see appendix~\ref{sec:appendix-stss}). The two plateaus in both $a(t)$ and $v(t)$ in figures~\ref{fig:polymer-simulation}(a--b), which arise for each parameter set, correspond to the conclusion of the initial poroelastic relaxation phase and the attainment of a steady state, if one exists, respectively.

For a fixed pressure drop, larger initial porosities lead to a larger final displacement of the left boundary since the solid skeleton is less dense in the material. This effect becomes more apparent at late times, as the material weakens (see figure~\ref{fig:polymer-simulation}(a)). If we instead fix the initial porosity of the material and increase the pressure drop, more fluid exits on the right of the domain which generates larger porosity gradients; this in turn pushes the solid skeleton towards the rigid boundary and increases the maximal value of $a$. As discussed in \S\ref{sec:qss-param-space}, if the pressure drop is too large and the initial porosity is too small, the porosity on the right of the domain drops to zero at a finite time, as is the case when $\phi_{f,0} = 0.5$ and $\gamma = 0.2$. Therefore, the pores close and fluid cannot flow through the whole domain; we terminate the simulation here. The numerical value of $\tcrit$ is $3.33$, whilst the analytical value (using (\ref{qss:t-crit})) is $3.25$. This error is within our numerical timestep (see appendix~\ref{sec:appendix-fenics}).

We study the case $(\phi_{f,0}, \gamma) = (0.5,0.2)$ further in figures~\ref{fig:polymer-simulation}(c--f). In the initial phase, between $t \sim 10^{-7}$ and $t \sim 10^{-3}$, the average porosity of the material decreases as the material compresses whilst the average solute concentration and Young's modulus remain constant (see figure~\ref{fig:polymer-simulation}(c)). The porosity instantly decreases on the right due to the solid bearing the compressive load from the fluid (figure~\ref{fig:polymer-simulation}(d)). In this simulation, $t_{\mathrm{pe}} / t_{\mathrm{weak}} = 3.0 \times 10^{-3}$ (see table~\ref{table:dless-parameter-vals}), and so the initial phase ends at around the poroelastic relaxation time. Before the second phase begins at approximately $t = 10^{-2}$, the fluid carries the solute from left to right (see figure~\ref{fig:polymer-simulation}(e)) until it reaches its full saturation within the pore space (figure~\ref{fig:polymer-simulation}(c)) at approximately the advection timescale ($t_{\mathrm{adv}} / t_{\mathrm{weak}} = 1.4\times 10^{-2}$). The left boundary position does not vary greatly in this period, meaning that advection does not play a substantial role in the deformation of the material. Solute advection dominates diffusion due to the large Péclet number of $1.6\times 10^{2}$.

At approximately $t = 10^{-2}$, $c$ has not quite reached its spatially uniform steady state but $E$ still decays. As a result, $E$ decays slightly more quickly near $x = a$ than near $x = 1$ at this time, which corresponds to the slight spatial gradient in $E$ across the domain (see inset of figure~\ref{fig:polymer-simulation}(f)). The softer material compresses further in the second phase of evolution, with $a$, $\phi_{f}$ and $v$ changing on the much slower timescale, $t_{\mathrm{weak}}$. To maintain the fixed pressure drop across the weakened material, the porosity also starts to decrease on the right. In this simulation, as $\phi_{f,0}$ is too small and $\gamma$ is too large, $\phi_{f,r}$ drops to zero and the pores close at a finite time (figure~\ref{fig:polymer-simulation}(d)), meaning this parameter set corresponds to case 2.

The results presented have shown that the model exhibits a variety of complex behaviours. In some situations, the weakening mechanism causes the pores to close at a finite time, and a steady state is not attained.

\section{Systems with comparable timescales} \label{sec:comparable-timescales}

We now relax the assumption that the weakening timescale is much longer than the poroelastic relaxation and advection timescales. Hence, (\ref{slow-weak:tE}) no longer holds. Instead, we assume that the solute behaves as a catalyst for the weakening process, so that it cannot be consumed upon interaction with the solid skeleton. In certain regions of $(\epsilon, \gamma, \Pen)$ parameter space, the Young's modulus of the material decays in a more heterogeneous manner compared to systems with slow weakening, controlled by the solute concentration profile. This in turn affects the value of the porosity on the right of the domain, and so the interplay between poroelastic deformation, weakening and solute transport are more intertwined than in \S\ref{sec:slow-weakening}. In addition, we investigate to what extent cases 1, 2 and 3 (described in \S\ref{sec:qss}) apply to systems without a separation of timescales.

\subsection{Numerical simulations example II: Dex-MA hydrogel}\label{sec:results-dexma}

In example II, we model the enzymatic weakening of a Dex-MA hydrogel in the presence of dextranase, which we assume to be dissolved in water flowing through the gel. We take $\phi_{f,0} = 0.5$, $\Emin = 0.7$ and $\gamma = 0.5$. From the work by~\citet{Meyvis1999}, we estimate a weakening timescale of $2.1 \times 10^{5}\ \mathrm{s}$ using figure 10 of their paper and a solute concentration scale of $4.1\times10^{-4}\ \mathrm{gL}^{-1}$. We multiply $c^{*}$ by factors of $0.1$, $1$, $10$ and $100$, with $\beta_E$ fixed, to analyse four simulations with different weakening timescales (see table~\ref{table:dless-parameter-vals}). We note that all four simulations start at the same dimensional time, but since time is scaled with $t_{\mathrm{weak}}$, each nondimensional start time is different.

\begin{figure*}
 \centering
 \includegraphics[height=15.5cm]{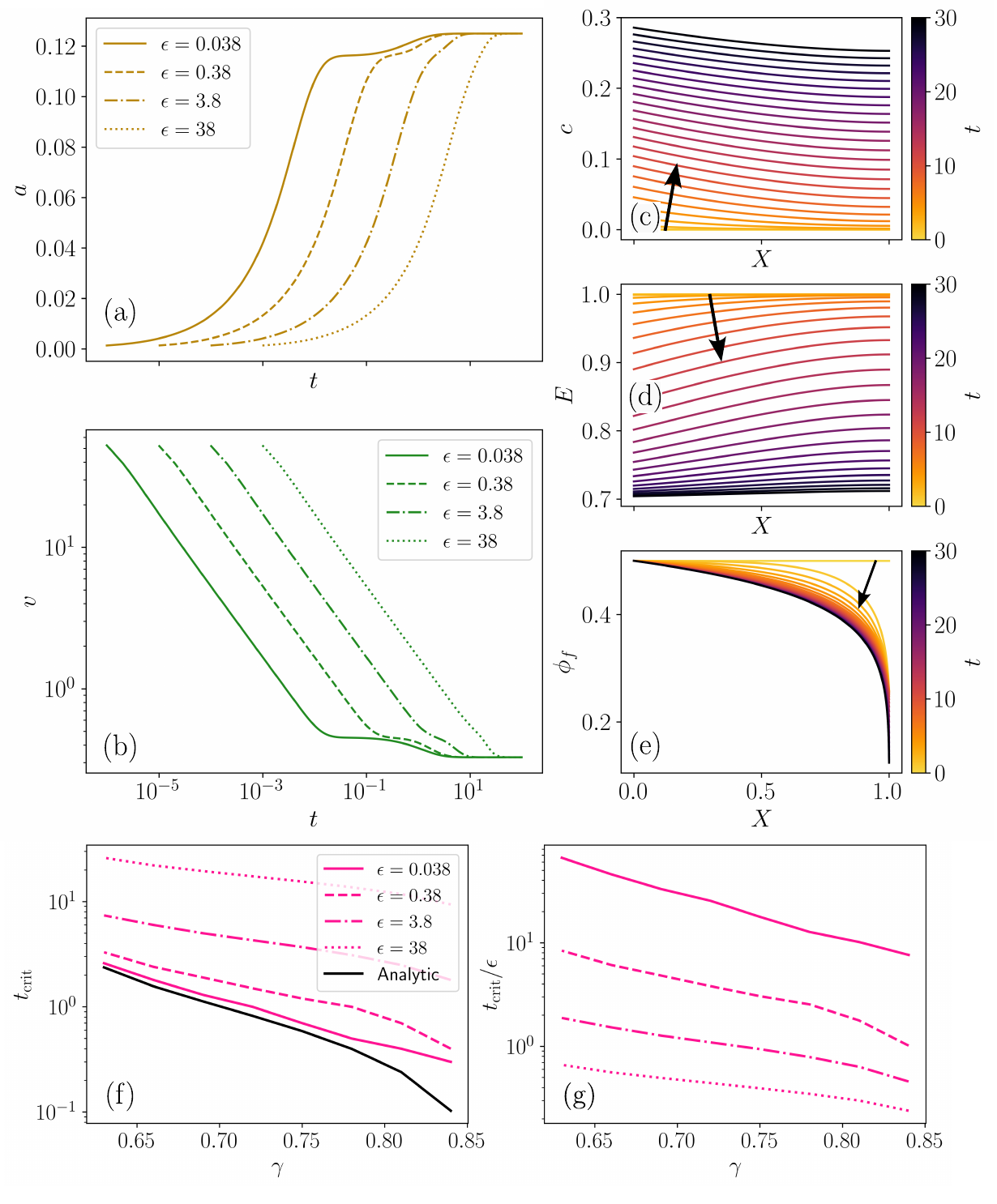}
 \caption{Numerical results for example II --- a Dex-MA hydrogel weakening in the presence of dextranase. The evolution of (a) the left boundary, $a$, and (b) the phase-averaged velocity, $v$, over time. We varied the value of $\epsilon$ across four different simulations. Next, we fix $\epsilon = 38$ and plot the spatial evolution of (c) the porosity $\phi_{f}$, (d) the Young's modulus $E$, and (e) the solute concentration $c$. The arrows denote the direction of increasing time. We plot against the Lagrangian coordinate, $X$. We plot the recorded values of (f) $\tcrit$ and (g) $\tcrit/\epsilon$ for given values of $\gamma$ and $\epsilon$ in case 2 simulations. For each value of $\epsilon$, we ran $10$ simulations, each with different values for $\gamma$, and recorded the time of pore closure, $\tcrit$. The black line in (f) represents the leading-order analytical $\tcrit$ given by (\ref{qss:t-crit}) for systems with slow weakening.}
 \label{fig:Dex-MA-simulation}
\end{figure*}

Small values of $c^{*}$ correspond to longer weakening timescales and smaller values of $\epsilon$, since there is less solute in the domain to weaken the material. When $\epsilon \ll 1$, as in \S\ref{sec:results-polymer}, the left boundary compresses inwards in two distinct phases, which gives rise to two separate plateaus (see solid line in figure~\ref{fig:Dex-MA-simulation}(a)). However, as we increase the value of $c^{*}$, the material weakens more quickly, and the earlier plateau disappears as the dynamics become more coupled. This transition happens as $\epsilon$ becomes $O(1)$ (see figure~\ref{fig:Dex-MA-simulation}(a)). Similarly, the early plateaus in the phase-averaged velocity disappear as $\epsilon$ is increased across simulations, as can be seen from figure~\ref{fig:Dex-MA-simulation}(b).

The increased coupling between the physical processes for the case $\epsilon = 38$ is apparent from the spatial evolution of the governing variables in figures~\ref{fig:Dex-MA-simulation}(c--e). Since $c^{*}$ is large, the material now weakens when $c$ is relatively small. The Péclet number for the solute is $0.16$ (see table~\ref{table:dless-parameter-vals}) meaning that this simulation is diffusion-dominated, though advection is the main driver of spatial variation in $c$ (see figure~\ref{fig:Dex-MA-simulation}(c)). Furthermore, the rate of decay of $E$ is proportional to solute concentration (see (\ref{nd:system-condensed-E})), and so the Young's modulus decays more quickly near $X = 0$ than near $X = 1$. Hence, the system exhibits spatially non-uniform weakening (see figure~\ref{fig:Dex-MA-simulation}(d), Movie 2). The heterogeneity in $c$, and hence $E$, would further increase if advection were more important (see figure~\ref{fig:polymer-simulation}(e)). As $\Emin$ is spatially uniform, the spatial heterogeneity in $E$ reduces as the Young's modulus approaches its steady state. Given that $t_{\mathrm{pe}} > t_{\mathrm{weak}}$, poroelastic relaxation occurs whilst the material is weakening, meaning that the porosity does not reach a quasi-steady state. Instead, the softer material on the left compresses into the stiffer material on the right, and the poroelastic relaxation and weakening phases combine (see figure~\ref{fig:Dex-MA-simulation}(e)).

The condition for existence of a steady state is determined by (\ref{ss:alpha}) and holds for all systems, regardless of the sizes of the fundamental timescales. In addition, if $\Emin < \alphamin \leq 1$ then the system is in case 2 and hence pore closure occurs after a finite nondimensional time $\tcrit$. Using our finite element implementation, we numerically compute $\gamma$ and $\epsilon$ values in the regime where case-2 dynamics occur and compare the results to the quasi-steady prediction given by (\ref{qss:t-crit}) (see figure~\ref{fig:Dex-MA-simulation}(f)). As we decrease the value of $\epsilon$ across simulations (or equivalently decrease the value of $c^{*}$), we approach the analytical value $\tcrit$, no matter the value of $\gamma$. For larger values of $\epsilon$, the Young's modulus weakens while $c$ is much less than its steady state value of $1$, since a smaller (dimensionless) concentration of solute is required to induce significant weakening. Hence, as $\epsilon$ increases, the rate of weakening and the time of pore closure become limited by solute transport and thus controlled by the advection and diffusion timescales, resulting in the dimensionless time of pore closure increasing with $\epsilon$. However, the dimensional time of pore closure, $t_{\mathrm{pe}}\tcrit/\epsilon$, decreases as $\epsilon$ increases when $t_{\mathrm{pe}}$ is fixed, since large values of $\epsilon$ correspond to large values of $c^{*}$ (see figure~\ref{fig:Dex-MA-simulation}(g)). Though the quasi-steady approximation to $\tcrit$ is not appropriate for all simulations, it provides a useful lower bound for the time of pore closure and is a good estimate for systems where $t_{\mathrm{weak}}$ is larger than the other timescales of the system.

\section{Conclusions}


We have developed a mathematical model for a nonlinear poroelastic material that weakens due to a solute or enzyme dissolved in the interstitial fluid. The model is used to examine flow-driven uniaxial compression of a weakening poroelastic material. Using a combination of asymptotic and numerical methods, we found that the behaviour of the material can be separated into three cases: the system reaches a steady state (case 1), the pores close after a finite amount of time (case 2), or the pores close immediately (case 3). In case 2, an analytical expression for the time of pore closure can be obtained in the limit of slow weakening. Case 2 is not possible for non-weakening poroelastic materials, subject to the kinematic and dynamic conditions we prescribe since the porosity on the right of the domain is fixed. In addition to this, we demonstrated that weakening the material decreases the maximal pressure drop that can be applied across the domain without closing the pores.


When the timescales of our model were sufficiently separated, as in example I (\S\ref{sec:results-polymer}), we found that the porosity evolved in two distinct phases: an early period dominated by poroelastic deformation, whereby the material compressed under fluid pressure; and a much later period over a longer timescale, in which the reduction in stiffness caused the material to compress further. In an experimental setting, if the left boundary were measured at different times, the later plateau could be used to estimate the weakening timescale for the system. In addition, if a time of pore closure were observed, the formulae derived could be used to estimate the minimum Young's modulus of the material.


In example II (\S\ref{sec:results-dexma}), the parameters for the Dex-MA/dextranase system gave rise to timescales of similar orders. When we modified the weakening timescale to be smaller than the poroelastic relaxation timescale, the solute concentration, and hence the weakening of the material, was spatially non-uniform. In addition, heterogeneity is more pronounced in advection-dominated flow of solute. In biomedical devices, this heterogeneity in stiffness could lead to internal points of material failure, and our mathematical model can be used to predict when and where this occurs. Furthermore, we found that the conditions which give rise to cases 1 -- 3 in systems with slow weakening apply to systems with comparable timescales, with the caveat that the time of pore closure cannot be estimated in the same way. The slow-weakening estimate for the time of pore closure relies on the assumption that the solute concentration has reached its steady-state value when the pores close, which is not necessarily the case when the timescales of the system are of similar orders. Crucially, we have classified model behaviour according to only the input parameters of the system; in future work this classification and the predicted value of $\tcrit$ could be experimentally validated.


Many biomedical applications have more intricate geometries than the one-dimensional channel considered in this work. For example, porous polymer scaffolds used in tissue engineering take on whichever shape is required to fulfil their role as synthetic extra-cellular matrix~\citep{Mi2014}. We specified the three-dimensional system in \S\ref{sec:3D-model}, which, as future work, could be used to investigate more complex geometries. Our mathematical model can also be extended to incorporate a broad range of additional physics. For example, although we model a poroelastic material, we could introduce a chemical potential into the model~\citep{Hong2008} to specialise to describe hydrogels, which are prone to hydrolytic degradation~\citep{Dhote2014,Pan2022}. We have also neglected the effect of solute clogging in pores, for example, via mineral precipitation~\citep{Yang2024}. Clogging would hamper the flow of fluid through the system and potentially reduce the deformation. Similarly, the tortuosity of a porous medium reduces the speed of diffusion of solute compared to that of a non-porous medium; hence, we could extend the model to include a porosity-dependent tortuosity term in the coefficient for molecular diffusion, $\D{m}$~\citep{Shen2007}. In addition, for large pore sizes, the flow through a porous medium is better described by the Darcy-Brinkman equation~\citep{Carrillo2019} which accounts for macroscale fluid shear effects that we do not consider here. We could also adapt our system to include a physical model of pore closure~\citep{Jannesari2026}, rather than letting the model break down. 


These extensions highlight the complexity of weakening poroelastic materials. The mechanistic insights gained through mathematical modelling are therefore expected to have broad impact across disciplines ranging from biomedicine to geoscience.

\bibliographystyle{jfm}
\bibliography{bibliographies/bibliography}


\backsection[Supplementary data]{\label{SupMat}Supplementary videos are available at (to be determined). Movie 1: animation of example I with $\phi_{f,0} = 0.6$ and $\gamma = 0.1$ (a case 1 simulation). Movie 2: animation of example II with $\epsilon = 38$.}

\backsection[Acknowledgements]{The authors would like to acknowledge Jessica Williams, Cristian Parisi, Niraj Rauniyar and Aditi Ray (all from Boston Scientific Corporation) for useful discussions.}

\backsection[Funding]{This work is partially funded by a research grant from Boston Scientific Corporation (BSC) and the Engineering and Physical Sciences Research Council (EPSRC)-funded Sustainable Approaches to Biomedical Science: Responsible and Reproducible Research Centre for Doctoral Training (SABS R\textsuperscript{3} CDT). MGH was supported by the EPSRC (Grant No. UKRI093). For the purpose of open access, the authors have applied a CC BY public copyright licence to any accepted manuscript arising from this submission.}

\backsection[Declaration of interests]{The authors report no conflict of interest.}

\backsection[Data availability statement]{The code used to generate the data that support the findings of this study are openly available in the following GitHub repository: \url{https://github.com/mghosh00/PoroelasticMaterials}.}

\backsection[Author ORCIDs]{M. V. Ghosh, https://orcid.org/0009-0001-9605-9888; M. G. Hennessy, https://orcid.org/0000-0002-5928-6256; A. Münch, https://orcid.org/0000-0002-8325-3809; S. L. Waters, https://orcid.org/0000-0001-5285-0523.}

\backsection[Author contributions]{CRediT contributions. M. V. Ghosh: Data curation, Formal analysis, Investigation, Methodology, Software, Validation, Visualisation, Writing --- original draft, Writing --- review and editing. M. G. Hennessy: Conceptualisation, Funding acquisition, Methodology, Software, Supervision, Writing --- review and editing. A. Münch: Conceptualisation, Funding acquisition, Methodology, Supervision, Writing --- review and editing. S. L. Waters: Conceptualisation, Funding acquisition, Methodology, Supervision, Writing --- review and editing.}

\appendix

\section{Derivation of the boundary and initial conditions of the solute concentration} \label{sec:appendix-c-ibcs}

\begin{figure*}
 \centering
 \includegraphics[height=7.5cm]{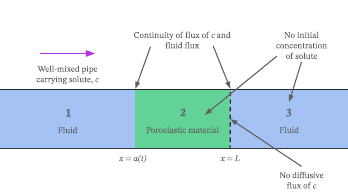}
 \caption{Diagram of the three regions considered in appendix~\ref{sec:appendix-c-ibcs}. Region 1 contains a well-mixed pipe of fluid carrying a solute. At time $t = 0$, the pipe is brought into contact with the poroelastic material in green in region 2, positioned between $x = a(t)$ and $x = L$. Region 3 also contains a bath of fluid, which is initially void of solute. We impose no diffusive flux of $c$ on the boundary between regions 2 and 3 and continuity conditions at both boundaries.}
 \label{fig:three-domains}
\end{figure*}

To calculate the boundary and initial conditions on $c$, we suppose that there is a pipe of well-mixed fluid to the left of the poroelastic material (region 1) carrying the dissolved solute, which begins to flow through the material at $t = 0$. Initially, there is no solute dissolved in region 2 (the poroelastic material) or region 3 (to the right of the material), as shown in figure~\ref{fig:three-domains}. Hence,

\begin{equation} \label{three-region:c-ic}
    c = 0 \qquad \mathrm{at} \quad t = 0, \quad a(t) < x < L.
\end{equation}

In region 1, as the fluid is well-mixed, we suppose that the concentration of solute is uniform throughout the pipe and satisfies $c = c^{*}$ for some constant $c^{*}$. From \S\ref{sec:3D-model}, we recall that in region 2, $c$ and $v_{f}$ satisfy

\vspace{-0.5cm}
\begin{subequations} \label{three-region:eqns-r2}
\begin{align}
\begin{split} \label{three-region:eqn-c-R2}
    \frac{\p}{\p t}(\phi_{f}c) + \frac{\p}{\p x}\left(\phi_{f}c v_{f} - \D{m}\phi_{f}\frac{\p c}{\p x}\right) &= 0,
\end{split}\\
\begin{split} \label{three-region:eqn-vf-R2}
    \frac{\p \phi_{f}}{\p t} + \frac{\p}{\p x}\left(\phi_{f}v_{f}\right) &= 0,
\end{split}
\end{align}
\end{subequations}

\noindent where $\phi_{f}$ is the porosity.

On the left interface between regions 1 and 2, we impose continuity of total solute flux and of fluid flux. Explicitly, using the conservation equations in (\ref{three-region:eqns-r2}), the following conditions must be satisfied:

\begin{equation} \label{three-region:cty-a}
    \left[\phi_{f}\left(\D{m}\frac{\p c}{\p x} - (v_{f}-\dot{a})c\right)\right]^{+}_{-} = 0, \qquad [\phi_{f}(v_{f} - \dot{a})]^{+}_{-} = 0 \qquad \mathrm{on} \quad x = a(t),
\end{equation}

\noindent where 

\begin{equation}
[f]^{+}_{-} := f\big|_{x_{0}^{+}} - f\big|_{x_{0}^{-}} \qquad \mathrm{on} \quad x = x_{0}.
\end{equation}

\noindent Following this, we combine conditions (\ref{three-region:cty-a}) with the fact that $c = c^{*}$ in region 1 to reach the following condition on $c$ in region 2:

\begin{equation} \label{three-region:c-a+}
    \phi_{f}\left(\D{m}\frac{\p c}{\p x} - (v_{f} - \dot{a})(c - c^{*})\right)\Bigg|_{a^{+}} = 0
\end{equation}

On the right interface, which separates regions 2 and 3, we suppose that there is no diffusive flux of solute through the boundary, so

\begin{equation} \label{three-region:c-L-}
    \phi_{f}\D{m}\frac{\p c}{\p x}\Bigg|_{L^{-}} = 0.
\end{equation}

\noindent Continuity of fluid flux and total solute flux must also be satisfied at this boundary, yet since fluid flows from left to right, the behaviour of the solute in region 2 dictates that of region 3. Hence, these continuity conditions can be used to determine the behaviour of the solute in region 3, though we do not provide an analysis of this here.

\section{Deriving the condensed one-dimensional model} \label{sec:appendix-deriving-condensed-model}

For uniaxial compression, the deformation gradient tensor is $\mathsfbi{F}=\mathrm{diag}(J, 1, 1)$. We use the form of the solid displacement in (\ref{1D:vec-to-scal}) with this expression for $\mathsfbi{F}$ to relate the Jacobian and the $x$-component of the solid displacement, $u_{s}$, in the following way:

\begin{equation} \label{deriv-cond:J-u_s}
    J = \frac{1}{1 - \frac{\p u_{s}}{\p x}}.
\end{equation}

\noindent Substituting (\ref{deriv-cond:J-u_s}) into (\ref{gen:solid-J-phi}), we can relate gradients in the solid displacement to the porosity:

\begin{equation} \label{deriv-cond:u_s}
    \frac{\p u_{s}}{\p x} = \frac{\phi_{f} - \phi_{f,0}}{1 - \phi_{f,0}}.
\end{equation}

\noindent Mass conservation for the solid skeleton (\ref{gen:solid-mass}) becomes simply

\begin{equation} \label{deriv-cond:solid-mass}
    \frac{\p \phi_{f}}{\p t} - \frac{\p}{\p x}\left[(1 - \phi_{f})v_{s}\right] = 0.
\end{equation}

\noindent Next, the fluid conservation of mass and momentum equations, from (\ref{gen:fluid-mass}) and (\ref{gen:fluid-darcy}), respectively, are

\vspace{-0.5cm}
\begin{subequations} \label{deriv-cond:solflu}
\begin{align}
\begin{split} \label{deriv-cond:fluid-mass}
    \frac{\p \phi_{f}}{\p t} + \frac{\p}{\p x}\left[\phi_{f}v_{f}\right] &= 0,
\end{split}\\
\begin{split} \label{deriv-cond:darcy}
    \phi_{f}(v_{f} - v_{s}) &= -\frac{k(\phi_{f})}{\mu}\frac{\p p_{f}}{\p x}.
\end{split}
\end{align}
\end{subequations}

\noindent Since the deformation gradient tensor is diagonal, from (\ref{gen:const-const}) we deduce that the Terzaghi stress tensor is also diagonal. Hence, we combine the equation for mechanical equilibrium (\ref{gen:mech-navier}) with (\ref{gen:const-sigma}) and take the $x$-component to get

\begin{equation} \label{deriv-cond:mech-eqm}
    \frac{\p\sigma_{xx}'}{\p x} = \frac{\p p_{f}}{\p x}.
\end{equation}

\noindent The three non-zero elements of the Terzaghi stress tensor are (found by inserting the diagonal tensor $\mathsfbi{F}$ into (\ref{gen:const-const}))

\vspace{-0.5cm}
\begin{subequations} \label{deriv-cond:terzaghi}
\begin{align}
\begin{split} \label{deriv-cond:sigma_xx'}
    \sigma_{xx}' &= \frac{E}{2(1 + \upnu)(1 - 2\upnu)J}\left[J^{2} - 2\upnu J - (1 - 2\upnu)\right],
\end{split}\\
\begin{split} \label{deriv-cond:sigma_yyzz'}
    \sigma_{yy}' &= \sigma_{zz}' = \frac{\upnu E}{(1 + \upnu)(1 - 2\upnu)}(J - 1),
\end{split}
\end{align}
\end{subequations}

\noindent where we define $\sigma_{yy}'$, $\sigma_{zz}'$ to be principal stresses in two directions orthogonal to the direction of deformation and fluid flow. We eliminate $J$ in favour of $\phi_{f}$ using (\ref{gen:solid-J-phi}) in (\ref{deriv-cond:sigma_xx'}) to determine that

\begin{equation} \label{deriv-cond:sigma_xx'-phi}
    \sigma_{xx}' = Eg(\phi_{f}).
\end{equation}

\noindent We note that $g$ encapsulates the neo-Hookean constitutive law for the material.

The solid and fluid velocities, $v_{s}$ and $v_{f}$, are written in terms of the phase-averaged velocity $v$ defined in (\ref{1D:v}) to find, using (\ref{deriv-cond:mech-eqm}) and (\ref{deriv-cond:sigma_xx'-phi}), that

\begin{equation} \label{deriv-cond:v_s-v-E-phi}
    v_{s} = v + \frac{k(\phi_{f})}{\mu}\frac{\p}{\p x}\left(E g(\phi_{f})\right),
\end{equation}

\noindent and

\begin{equation} \label{deriv-cond:v_f-v-E-phi}
    v_{f} = v - \frac{1 - \phi_{f}}{\phi_{f}}\frac{k(\phi_{f})}{\mu}\frac{\p}{\p x}\left(Eg(\phi_{f})\right).
\end{equation}

\noindent We then subtract (\ref{deriv-cond:solid-mass}) from (\ref{deriv-cond:fluid-mass}) and insert (\ref{1D:v}) into the resulting equation; this yields

\begin{equation} \label{deriv-cond:div-v}
    \frac{\p v}{\p x} = 0,
\end{equation}

\noindent which represents mass conservation for the poroelastic material. We insert the expression (\ref{deriv-cond:v_s-v-E-phi}) into the equation for mass conservation of the solid (\ref{deriv-cond:solid-mass}) and use (\ref{deriv-cond:div-v}) to get the following nonlinear advection-diffusion equation for the porosity:

\begin{equation} \label{deriv-cond:phi-diffusion}
    \frac{\p \phi_{f}}{\p t} + v\frac{\p \phi_{f}}{\p x} = \frac{\p}{\p x}\left[\frac{(1 - \phi_{f})k(\phi_{f})}{\mu}\frac{\p}{\p x}\left(Eg(\phi_{f})\right)\right].
\end{equation}

The solute transport equation (\ref{gen:weak-c}) reduces to

\begin{equation} \label{deriv-cond:c}
    \frac{\p}{\p t}(\phi_{f}c) = \frac{\p}{\p x}\left(\D{m}\phi_{f}\frac{\p c}{\p x} - \phi_{f}c v_{f}\right).
\end{equation}

\noindent We insert the equation for mass conservation of the fluid (\ref{deriv-cond:fluid-mass}) into the equation for $c$ (\ref{deriv-cond:c}), which leads to

\begin{equation} \label{deriv-cond:c-simplified}
    \phi_{f}\left(\frac{\p c}{\p t} + v_{f}\frac{\p c}{\p x}\right) = \frac{\p}{\p x}\left(\D{m}\phi_{f}\frac{\p c}{\p x}\right).
\end{equation}

\noindent In summary, the four condensed governing equations are

\vspace{-0.5cm}
\begin{subequations} \label{deriv-cond:system-condensed}
\begin{align}
\begin{split} \label{deriv-cond:system-condensed-phi}
    \frac{\mathrm{D}\phi_{f}}{\mathrm{D}t} &= \frac{\p}{\p x}\left[\frac{(1 - \phi_{f})k(\phi_{f})}{\mu}\frac{\p}{\p x}\left(Eg(\phi_{f})\right)\right],
\end{split}\\
\begin{split} \label{deriv-cond:system-condensed-v}
    \frac{\p v}{\p x} &= 0,
\end{split}\\
\begin{split} \label{deriv-cond:system-condensed-E}
    \frac{\mathrm{D}^{s}E}{\mathrm{D}t} &= -\frac{1}{t_{\mathrm{weak}}c^{*}}c\left(E - \Emin\right),
\end{split}\\
\begin{split} \label{deriv-cond:system-condensed-c}
    \phi_{f}\frac{\mathrm{D}^{f}c}{\mathrm{D}t} &= \frac{\p}{\p x}\left(\D{m}\phi_{f}\frac{\p c}{\p x}\right),
\end{split}
\end{align}
\end{subequations}

\noindent where we define

\vspace{-0.5cm}
\begin{subequations} \label{deriv-cond:mater-deriv}
\begin{align}
\begin{split} \label{deriv-cond:D_Dt}
    \frac{\mathrm{D}}{\mathrm{D}t} &:= \frac{\p}{\p t} + v\frac{\p}{\p x},
\end{split}\\
\begin{split} \label{deriv-cond:Ds_Dt}
    \frac{\mathrm{D}^{s}}{\mathrm{D}t} &:= \frac{\p}{\p t} + v_{s}\frac{\p}{\p x},
\end{split}\\
\begin{split} \label{deriv-cond:Df_Dt}
    \frac{\mathrm{D}^{f}}{\mathrm{D}t} &:= \frac{\p}{\p t} + v_{f}\frac{\p}{\p x},
\end{split}
\end{align}
\end{subequations}

\noindent and $v_{s}$ and $v_{f}$ are given by (\ref{deriv-cond:v_s-v-E-phi}) and (\ref{deriv-cond:v_f-v-E-phi}), respectively. These are only algebraic relations and do not require solving further differential equations.

The initial conditions for $\phi_{f}$, $E$ and $c$, together with that for the boundary position $a(t)$, are given in (\ref{1D:ICs}). The boundary conditions for $c$ are given in (\ref{1D:BCs-c}). It remains to convert the kinematic and dynamic boundary conditions (\ref{1D:BCs-kin})--(\ref{1D:BCs-dyn}) into boundary conditions on the porosity and equations for determining the left boundary and phase-averaged velocity. Firstly, we use (\ref{nd:g}) to rewrite the conditions on the Terzaghi stress (\ref{1D:BCs-dyn}) in terms of the porosity and Young's modulus in the following way:

\vspace{-0.5cm}
\begin{subequations} \label{deriv-cond:BCs-phi}
\begin{align}
\begin{split} \label{deriv-cond:BCs-phi-a}
    \phi_{f} = \phi_{f,0} \qquad &\mathrm{on} \quad x = a(t),
\end{split}\\
\begin{split} \label{deriv-cond:BCs-phi-1}
    Eg(\phi_{f}) = -\Delta p \qquad &\mathrm{on} \quad x = L.
\end{split}
\end{align}
\end{subequations}

\noindent Condition (\ref{deriv-cond:BCs-phi-a}) arises since we impose no solid stress on the free boundary meaning the elastic strain is zero at $x = a$, whilst condition (\ref{deriv-cond:BCs-phi-1}) implies that all load is carried by the solid at $x = L$. Condition (\ref{deriv-cond:BCs-phi-a}) is a consequence of the fact that $g(\phi_{f})$ has a unique root in the interval $(0, 1)$ at $\phi_{f} = \phi_{f,0}$.

From (\ref{deriv-cond:system-condensed-v}) we see that the phase-averaged velocity $v$ is spatially uniform; thus, we can write $v = v(t)$. To determine $v(t)$, we substitute expression (\ref{deriv-cond:v_s-v-E-phi}) into the kinematic condition on $v_{s}$ in (\ref{1D:BCs-kin-L}) to get

\begin{equation} \label{deriv-cond:constraint-v}
    v(t) = -\frac{k(\phi_{f})}{\mu}\frac{\p}{\p x}\left(Eg(\phi_{f})\right)\Bigg|_{x=L}.
\end{equation}

\noindent We integrate (\ref{deriv-cond:u_s}) from $a(t)$ to $L$ and insert the boundary conditions on the displacement in (\ref{1D:BCs-u_s}) to get the following implicit solution for $a(t)$:

\begin{equation} \label{deriv-cond:constraint-a}
    a(t) = \phi_{f,0}L - \int_{a(t)}^{L}\phi_{f}(x, t)\mathrm{d}x.
\end{equation}

\noindent System (\ref{deriv-cond:system-condensed}), initial conditions (\ref{1D:ICs}), boundary conditions (\ref{1D:BCs-c}) and (\ref{deriv-cond:BCs-phi}), and equations (\ref{deriv-cond:constraint-v})--({\ref{deriv-cond:constraint-a}}) fully specify the one-dimensional problem.

\section{Details of the simulation and finite-element set-up} \label{sec:appendix-fenics}

We use the Python package FEniCS~\citep{Logg2012,Alnaes2015} to conduct finite element simulations of the nondimensional model described in \S\ref{sec:nondimensional}. The code for simulating the model is in the following repository: \url{https://github.com/mghosh00/PoroelasticMaterials}. For each simulation, we use an equally spaced mesh with $1001$ grid points in $\xi$ space, where $\xi = (x - a) / (1 - a)$. We implement a finite difference method with implicit time stepping to resolve the time derivatives, whereby we use an exponential-spaced timestep in example I and a constant timestep in example II. In example I, the timepoints are given by the formula $t = 2\times10^{-6}\left(e^{0.025\tau}-1\right)$, for integers $\tau\in\{0,...,650\}$. In the simulation with pore closure ($\phi_{f,0} = 0.5$, $\gamma = 0.2$), $\tau$ lies between $0$ and $573$. In example II, the timepoints are $t = 0.1\tau$ ($0\leq\tau\leq300$). We use $P_1$ finite elements to represent the porosity, Young's modulus, solute concentration, displacement, fluid pressure, and Terzaghi stress at each time step. The left boundary and phase-averaged velocity are treated as spatially uniform Lagrange multipliers. To evaluate the solution at the first time point, we use the method described in appendix~\ref{sec:appendix-stss} to find a similarity solution. We solve the full problem using a nonlinear variational solver (with the similarity solution as an initial condition) for all other timepoints.

To determine the phase-averaged velocity, we integrate the following equation across the spatial domain:

\begin{equation} \label{fem:v}
    v = v_{s} - \frac{\gamma^{-1}}{1 - a}k(\phi_{f})\frac{\p}{\p \xi}\left(Eg(\phi_{f})\right),
\end{equation}

\noindent where we set

\begin{equation} \label{fem:vs}
    v_{s} = \epsilon\gamma^{-1}\frac{1}{1 - \phi_{f}}\left((1 - \phi_{f,0})\frac{\p u_{s}}{\p t} - (1 - \xi)\dot{a}(\phi_{f} - \phi_{f,0})\right),
\end{equation}

\noindent and $u_{s}$ is the nondimensional solid displacement. Equation (\ref{fem:vs}) comes from using the solid velocity, $v_{s} = \epsilon\gamma^{-1}\p u_{s} / \p t + v_{s}\p u_{s} / \p x$, and the relation $\p u_{s} / \p x = (\phi_{f} - \phi_{f,0}) / (1 - \phi_{f,0})$. 

The averages $\overline{\phi_{f}}$, $\overline{E}$ and $\overline{c}$ are computed as a post-processing step. If at any point the porosity drops below zero, we terminate the simulation. For plotting, we use the profile $u_{s}$ at each timepoint to determine the other quantities in terms of the Lagrangian coordinate, $X$.

\section{Small-time similarity solution} \label{sec:appendix-stss}

For the initial timepoint, we solve the early-time similarity solution problem, assuming the initial timestep is sufficiently small such that $E \approx 1$. In this problem, we set $\phi_{f} = \phi_{f,0} - f(\eta)$, where $\eta = (1 - x) / \sqrt{t}$, so $g = g\left(\phi_{f,0} - f(\eta)\right)$. We also set $v = C_{v}/\sqrt{t}$ and $a = C_{a}\sqrt{t}$ for constants $C_{v}$ and $C_{a}$. The second-order ordinary differential equation for the variable $f$ in the limit $t \rightarrow 0$ is

\begin{equation} \label{stss:f}
    \left(\gamma C_{v}+\frac{1}{2}\epsilon\eta\right)\frac{\mathrm{d}f}{\mathrm{d}\eta} = \frac{\mathrm{d}}{\mathrm{d}\eta}\left[\left(f+1-\phi_{f,0}\right)k(f)\frac{\mathrm{d}g}{\mathrm{d}\eta}\right],
\end{equation}

\noindent subject to the boundary conditions

\begin{equation} \label{stss:bcs}
    g = -\gamma \qquad \mathrm{at} \quad \eta = 0, \qquad f \rightarrow 0 \qquad \mathrm{as} \quad \eta \rightarrow \infty.
\end{equation}

\noindent The equation for determining the constant $C_{a}$ is

\begin{equation} \label{stss:a}
    C_{a} = \frac{1}{1-\phi_{f,0}}\int_{0}^{\infty}f(\eta)\mathrm{d}\eta.
\end{equation}

\noindent We couple this system with a similarity equation for the solid displacement, $u_{s}$, as this is needed to eliminate the constant $C_{v}$ from (\ref{stss:f}). However, we omit this equation here for brevity.

In the limit $\gamma \rightarrow 0$, which is equivalent to linear poroelasticity, this system admits leading-order analytical solutions. We expand $f \sim \gamma f^{(0)} + \gamma^{2} f^{(1)} + \cdots$, $C_{a} \sim C_{a}^{(0)} + \gamma C_{a}^{(1)} + \cdots$ and $C_{v} \sim C_{v}^{(0)} + \gamma C_{v}^{(1)} + \cdots$ as $\gamma \rightarrow 0$, and deduce the following leading-order solutions of (\ref{stss:f})--(\ref{stss:a}):

\vspace{-0.5cm}
\begin{subequations} \label{stss:lo-analytical}
\begin{align}
\begin{split} \label{stss:f0}
    f^{(0)}(\eta) &= \frac{1-\phi_{f,0}}{\D{\phi}}\mathrm{erfc}\left(\frac{\eta}{2\sqrt{\D{\phi}}}\right),
\end{split}\\
\begin{split} \label{stss:Ca0-Cv0}
    C_{a}^{(0)} = &\frac{2}{\sqrt{\pi\D{\phi}}}, \qquad C_{v}^{(0)} = \frac{1}{\sqrt{\pi\D{\phi}}},
\end{split}
\end{align}
\end{subequations}

\noindent where $\D{\phi}:=(1-\upnu)/\left((1+\upnu)(1-2\upnu)\right)$.


\section{Cross-validation of small-time, large-time and numerical solutions}\label{sec:appendix-validation}

\begin{figure*}
 \centering
 \includegraphics[height=8cm]{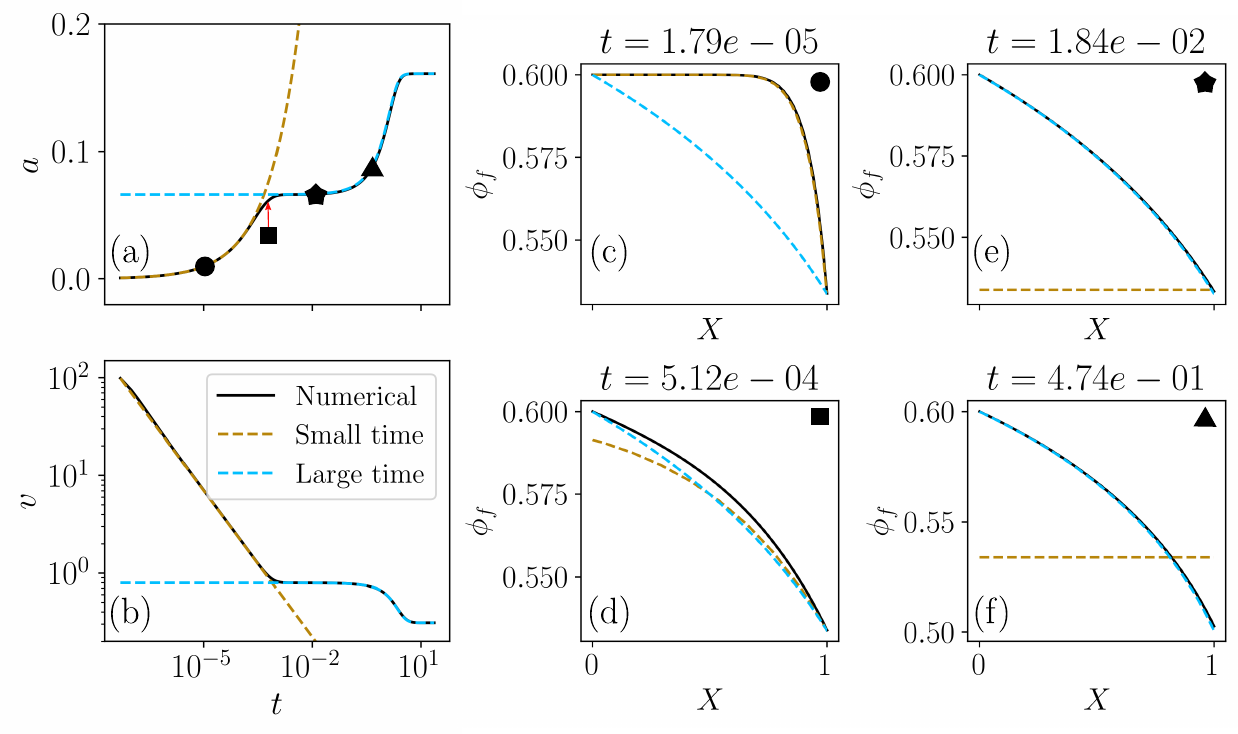}
 \caption{Comparison of numerical and leading-order analytical profiles in example I. Solid black represents the numerical solution, dashed gold is the small-time similarity solution, and dashed blue is the quasi-steady solution. For this simulation, we fixed $\phi_{f,0} = 0.6$ and $\gamma = 0.2$. We plot (a) the left boundary and (b) the phase-averaged velocity over time in each case. We also plot snapshots of the porosity at four timepoints in (c--f). Each shape in the top-right of panels (c--f) corresponds to the time in which the porosity is recorded in panel (a).}
 \label{fig:validation}
\end{figure*}

As validation for both the analytical approximations and the finite element simulations, we compare porosity snapshots, left boundary position and phase-averaged velocity for example I of the various solutions in figure~\ref{fig:validation}. In this simulation, $\gamma = 0.2$, $\phi_{f,0}=0.6$ and $\epsilon = 2.8\times10^{-3}$; see table~\ref{table:dless-parameter-vals} for other parameter values. Specifically, we solve the system (\ref{stss:f})--(\ref{stss:a}) to plot the small-time similarity solution and we use the solution (\ref{qss:phi-soln}) for the large-time quasi-steady solution of systems with slow weakening.

The small-time similarity solution fits well to the numerical solution up until approximately $t = 10^{-4}$ for both the left boundary and phase-averaged velocity (see figures~\ref{fig:validation}(a--b)). The porosity also fits well in this period (figure~\ref{fig:validation}(c)). There is a short transition period between $t = 10^{-4}$ and $t = 10^{-3}$ in which the small-time solution is inappropriate, as $a$ is no longer proportional to $\sqrt{t}$, but also the quasi-steady solution is inappropriate, since the system has not yet reached the initial poroelastic relaxation (see figure~\ref{fig:validation}(d)). Once this is reached, the large-time solution matches the numerical solution well (figure~\ref{fig:validation}(e)); the solutions continue to match as the material weakens (figure~\ref{fig:validation}(f)). The slight discrepancy between the two solutions at the right of the domain is a consequence of $E$ not being entirely spatially uniform.

\end{document}